\documentstyle[graphicx,cite]{mn}
\bibdata{refs}
\bibstyle{mnras3}

\newcommand{\ls}
 {\mathrel{\hbox{\rlap{\hbox{\lower4pt\hbox{$\sim$}}}\hbox{$<$}}}}
\newcommand{\gs}       
 {\mathrel{\hbox{\rlap{\hbox{\lower4pt\hbox{$\sim$}}}\hbox{$>$}}}}
\newcommand{\degrees}{\hbox{$^\circ$}}

\newcommand{\exosat}{{\it EXOSAT}}
\newcommand{\einstein}{{\it Einstein}}
\newcommand{\rosat}{{\it ROSAT}}

\newcommand{\heao}{{\it HEAO-1}}
\newcommand{\asca}{{\it ASCA}}
\newcommand{\ginga}{{\it GINGA}}

\newcommand{\chandra}{{\it Chandra}}

\title[ASCA observations of non-magnetic cataclysmic variables]
{The complete set of ASCA X-ray observations of non-magnetic cataclysmic 
variables}

\author[D.S.\ Baskill, P.J.\ Wheatley and J.P.\ Osborne]
        {Darren S.\ Baskill, Peter J.\ Wheatley and Julian P.\ Osborne\\
Department of Physics and Astronomy, University of Leicester, Leicester LE1 7RH, U.K.\\
}

\begin{document}

\maketitle

\label{firstpage}

\begin{abstract}
We present the complete set of thirty four \asca\ observations of
non-magnetic cataclysmic variables. 
Timing analysis reveals large
X-ray flux variations in dwarf novae in outburst (Z~Cam, SS~Cyg and
SU~UMa) and orbital modulation in high inclination systems (including OY~Car,
HT~Cas, U~Gem, T~Leo).
We also found episodes of unusually low accretion rate during quiescence (VW~Hyi and SS~Cyg).
Spectral analysis reveals broad temperature distributions in
individual systems, with emission weighted to lower
temperatures in dwarf novae in outburst. Absorption in excess of
interstellar values is required in dwarf novae in outburst, but not in
quiescence. We also find evidence for sub-solar abundances and X-ray
reflection in the brightest systems.

LS~Peg, V426~Oph and EI~UMa have X-ray spectra that are 
distinct from the rest of the sample
and all three exhibit candidate X-ray periodicities. We argue that they
should be reclassified as intermediate polars. 

In the case of V345~Pav we found that the X-ray source had been previously misidentified.

\end{abstract}


\section{Introduction}
Cataclysmic variables are close binary stars in which a
white dwarf accretes material lost from a Roche-lobe filling late-type 
companion. 
Non-magnetic cataclysmic variables are the subset of systems in which
the magnetic field of the white dwarf is not sufficiently strong to
dominate the dynamics of the accretion flow. In these systems the
accreting material forms an accretion disc around the white dwarf (see
\ncite{warnersbook} for a review of cataclysmic variables). 

The proximity of many cataclysmic variables and the relative
faintness of their stellar components makes their accretion discs
unusually accessible. Their discs also evolve on accessible 
timescales of days
and weeks, allowing us to make detailed observations of their behaviour. 
An important subclass of non-magnetic cataclysmic variables are the
dwarf novae, so named because of dramatic outbursts in which their
accretion discs brighten by a factor of around one hundred. These
outbursts are 
believed to be the result of a thermal-viscous instability in the
accretion disc (for a review see \ncite{lasotareview}). 

Non-magnetic cataclysmic variables tend to be reasonably bright X-ray
sources with luminosities of $10^{30}$--$10^{32}$\,erg\,s$^{-1}$ 
\egcite{rosatcvs}. Observations of eclipsing systems have shown that
X-ray emission originates very close to the white dwarf, 
at least in quiescence \egcite{htcas,zcha,oycareph,oycarxmmpj}. 
X-rays are probably emitted through shock
heating of gas in a narrow boundary layer
between the accretion disc and white dwarf where the accreting
material gives up its kinetic energy and settles onto the white dwarf
surface. X-ray observations of non-magnetic cataclysmic variables
are therefore
sensitive to the accretion rate through the accretion disc, and to the
conditions in the inner accretion disc. 

In this paper we present the complete set of \asca\ X-ray observations
of non-magnetic cataclysmic variables. 
Previous work on samples of X-ray observations of non-magnetic
cataclysmic variables has included observations from  
\heao\  \cite{heao1}, 
\einstein\ \cite{einsteincvsample,powerspeceinstein,spectraeinstein}, 
\exosat\  \cite{exosatME}, 
and \rosat\  \cite{Vrtilek94,10cvs,rosatnmcvs,richman96,rosatcvs}. 
\asca\  made long observations of cataclysmic
variables, with a high effective area across a broad energy range and higher spectral resolution than these previous instruments.

\section{Observations}

\subsection{The \asca\ Observatory}

The Japanese X-ray observatory \asca\ (\ncite{asca}) operated for
eight years between 1993 February 20 and 2000 July 14.
Its science instruments consisted of four identical
grazing-incidence X-ray telescopes 
(\ncite{ascaxrt}).  
These 
telescopes provided sufficient imaging resolution 
for background rejection and sufficient effective area to detect
cataclysmic variables up to  10\,keV. 

\asca\ was notable for being the first X-ray observatory to carry CCD 
detectors, providing excellent spectral resolution  (2 per cent at 5.9\,keV). 
Two of the X-ray telescopes were fitted with identical CCD cameras,
the Solid-state Imaging Spectrometers (SIS0 and SIS1). 
The other two telescopes were fitted with Gas Imaging Spectrometers
(GIS2 and GIS3), which had less spectral resolution by a factor
$\sim$4, but which were more sensitive above 5\,keV and had a much larger
field of view. Some of the detections discussed in this paper were
serendipitous and made only with the GIS detectors. 
Both types of detectors were sensitive in the range $\sim$0.6--10\,keV.

\subsection{Data Reduction}
\label{sec:reduction}

\subsubsection{Data screening}
\label{sec:screening}

Data from \asca\ were provided in a relatively raw form and required
screening to remove intervals of high background. A standard screening
recipe is provided by NASA\footnote{http://heasarc.gsfc.nasa.gov/docs/asca/abc/abc.html},
but since many of our targets were faint we decided to maximise our 
coverage by applying our own screening criteria. This
involved an iterative process in which the screening parameters for
each observation were progressively tightened until we were confident
that no high-background intervals remained. As a consequence, different
screening parameters were applied to each observation. In a few cases
individual high-background intervals were excluded by hand. 

The parameters we adjusted between observations were: the maximum
allowed telescope deviation from the source position (in the range
0.01--0.02\degrees), the minimum allowed angle between the spacecraft pointing 
and the limb of the Earth (5--18\degrees), 
and the minimum allowed angle between the spacecraft pointing
and the sunlit limb of the Earth (for SIS only; 5--35\degrees).
By selecting different values of these parameters for each observation
we were able to rescue many hours of coverage of sources which
would otherwise have been lost due to minor excursions across the
standard screening limits. 

For all the observations we also applied a limit to the background
monitor count rate of RBM\_CONT$<250$ 
(compared with RBM\_CONT$<100$ in the standard screening). 
We did not screen on cut-off rigidity (COR). All other
parameters were held at the standard screening values.

\begin{table*}
\caption{Log of \asca\ observations of non-magnetic cataclysmic variables.  
Types are taken from \protect\scite{ritter98} and are defined as 
follows: DN=dwarf nova, SU=SU~UMa subtype, UG=U~Gem subtype, 
ZC=Z~Cam subtype, NL=nova-like, Na=fast decline nova, AC=AM~CVn 
star consisting of two He white dwarfs, SH=optical superhumps 
have been observed, UX=UX~UMa subtype (permanent outburst), 
VY=VY~Scl subtype (anti-dwarf nova).   
The optical state refers to the outburst state of the disc during the \asca\ 
observation, as estimated from AAVSO online light curves, 
where OB=outburst, T=transition, Q=quiescence, 
and HS=high state for the nova-like variables.
Information on the optical state is not available for all systems. 
In the count rate column, $<$ indicates a non-detection.
The GIS2 hardness are the ratio of counts above and below 2.3\,keV.  \newline
References to previous publications of these data are (numbered) as follows: 
1) \protect\scite{Mauche02}; 
2) \protect\scite{baskill01}; 
3) \protect\scite{htcas}; 
4) \protect\scite{ascasscyg}; 
5) \protect\scite{do97}; 
6) \protect\scite{ugemszkody}; 
7) \protect\scite{tleoasca}; 
8) \protect\scite{wzsgeasca}.\newline
Distance references (letters) are as follows:
a) \protect\scite{hippdist},
b) \protect\scite{thorstensendistances};
c) \protect\scite{Harrison00};
d) \protect\scite{Harrison03};
e) \protect\scite{berriman85};
f) \protect\scite{patterson84};
g) \protect\scite{crboo01};
h) \protect\scite{oycarbruch96};
i) \protect\scite{wood95};
j) \protect\scite{v436cen};
k) \protect\scite{sproats};
l) \protect\scite{vwhyidist};
m) \protect\scite{warner87};
n) \protect\scite{v345pav};
p) \protect\scite{ktperdist};
q) \protect\scite{cppupdist}. \newline
Inclination references (Greek letters) are as follows:
$\alpha$) \protect\scite{v603aql2000};
$\beta$) \protect\scite{shaftersthesis};
$\gamma$) \protect\scite{ritter98};
$\delta$) \protect\scite{crboo01};
$\varepsilon$) \protect\scite{oycar89};
$\zeta$)  \protect\scite{htcasinc};
$\eta$) \protect\scite{wwcet4};
$\theta$) \protect\scite{alcom98};
$\iota$) \protect\scite{thesis};
$\kappa$) \protect\scite{gpcom99};
$\lambda$) \protect\scite{22cvs};
$\mu$) \protect\scite{ugeminc};
$\nu$) \protect\scite{vwhyiinc};
$\xi$) \protect\scite{tleoinc};
$\pi$) \protect\scite{bklyninc};
$\rho$) \protect\scite{v426oph88};
$\sigma$) \protect\scite{v345pav};
$\tau$) \protect\scite{lspeginc};
$\upsilon$) \protect\scite{cppupinc};
$\phi$) \protect\scite{wzsgeinc};
$\chi$) \protect\scite{suumainc};
$\psi$) \protect\scite{cuvel};
$\omega$) \protect\scite{ixvel}. 
}
\label{table:general}
\begin{center}
\begin{tabular}{l@{\hspace{1.5mm}}l@{\hspace{1.5mm}}c@{\hspace{1.5mm}}c@{\hspace{1.5mm}}c@{\hspace{2.5mm}}c@{\hspace{2.5mm}}c@{\hspace{1.5mm}}c@{\hspace{2.5mm}}c@{\hspace{1.5mm}}c@{\hspace{0.5mm}}c@{\hspace{0.5mm}}c@{\hspace{1.5mm}}c}
 &  & Orbital &  & & & Obs. & Exp. &  \multicolumn{2}{c}{Count rate} &  Hardness & & \\
 & & Period & & Observation &Optical & length & Time & \multicolumn{2}{c}{[s$^{-1}$]}  & ratio & Distance & \\
 Target & Type & (hrs) & Inclination & Date & State & [ks]& [ks]& SIS0 & GIS2 & GIS2 & [pc] & \footnotesize References \normalsize \\
\hline
V603 Aql & Na SH & 3.314 & 13$^o\pm$2$^o$ & 1996 Oct 8 & HS & 115  & 45  & 0.81 & 0.44 & 0.67 &237$\pm^{380}_{\phantom{3}90}$ & a,$\alpha$\\
TT Ari & NL VY & 3.301 & -& 1994 Jan 20 & HS & 50  & 20  & 0.38 & 0.23 & 0.69& $>$180 & 1,e \\
KR Aur & NL VY & 3.907 & 38$^o\pm$10$^o$& 1996 Mar 6 & HS & 41  & 22  & 0.07 & 0.03 &0.58& 180& 1,f,$\beta,\gamma$ \\
CR Boo & NL AC& 0.409 & 30$^o\pm$5$^o$& 1999 Jan 6 & HS?& 95  & 43  & 0.03 & 0.02 &0.35& 450$\pm$50 & g,$\delta$\\
Z Cam & DN ZC & 6.956 & 57$^o\pm$11$^o$& 1995 Mar 7 & OB & 353  & 80  & 0.04 & 0.02 &0.59& 163$^{+68}_{-38}$ & 2,b,$\beta,\gamma$\\
 & & & & 1997 Apr 12 & T & 36  & 17  & 0.39 & 0.26 &1.18& & \\ 
OY Car & DN SU & 1.515 & 83.3$^o\pm$0.2$^o$& 2000 Jan 28 & Q & 150  & 42  & 0.04 & 0.03 &1.03&86$\pm$4 & h,$\epsilon$\\
HT Cas & DN SU & 1.768 & 81$^o\pm$1$^o$ &1994 Sep 6 & Q & 82  & 34  & 0.07 & 0.05& 1.00& 165$\pm$16 & 3,i,$\zeta$\\
V436 Cen & DN SU & 1.500 & 65$^o\pm$5$^o$ &1997 Jun 19 & Q & 66  & 24  & 0.44 & 0.27 &0.78& 263 & j,$\gamma$\\ 
WW Cet & DN & 4.219 & 54$^o\pm$4$^o$& 1996 Dec 24 & Q & 41  & 24  & 0.44 & 0.25 &0.59& 121-171 & k,$\eta,\gamma$\\
AL Com$\dagger*$ & DN SU & 1.360 & 20$^o$-40$^o$ &1997 Jun 22 & ? & 101  & 42  & - & $<$0.005 & -&187-264 &k,$\theta$\\
GO Com$\dagger$ & DN SU & 1.579&  -&1996 Jan 8 & ?& 55  & 23  & - & 0.01 &0.95& 361-510 & k\\
GP Com & NL AC & 0.775 & 49$^o$-70$^o$& 1994 Jul 6 & ? & 41  & 7  & 0.13 & 0.07 &0.46& 68$^{+7}_{-6}$ & b,$\iota,\kappa$\\
 & & & & 1994 Jun 30 & ? & 93  & 27  & 0.20 & 0.09 & 0.52& & \\
EY Cyg$\dagger$ & DN UG & 5.244 &  $<$60$^o$&1999 Oct 23 & Q & 84  & 31  & - & 0.02 & 0.89&- & $\lambda$\\
SS Cyg & DN UG & 6.603 &  37$^o\pm$5$^o$&1993 May 26 & OB & 78  & 31  & 1.46 & 0.67&0.53&166$\pm13$&4,5,c,$\beta,\gamma$\\
 & & &  &1995 Nov 27 & Q & 46  & 23  & 0.53 & 0.30&0.68& & \\
U Gem & DN UG & 4.246 & 69.7$^o\pm$0.7$^o$& 1994 Oct 20 & Q & 71  & 34  & 0.35 & 0.21&0.73&96$\pm^{5}_{4}$&6,c,$\nu$\\
 & & &  & 1994 Nov 11 & Q & 59  & 29  & 0.38 & 0.22&0.78&&\\
VW Hyi & DN SU & 1.783 & 60$^o\pm$10$^o$& 1993 Nov 8 & Q & 37  & 14  & 0.12 & 0.08 &0.44& 82$\pm$5 &l,$\mu$\\
 & & & &1995 Mar 6 & Q & 27  & 6  & 0.03 & 0.02& 0.51&& \\
T Leo & DN SU & 1.412 & 28$^o$-65$^o$& 1998 Dec 13 & Q & 72  & 31  & 0.30 & 0.17&0.51&101$^{+13}_{-11}$&7,b,$\xi$\\
BK Lyn$\dagger*$ & NL SH & 1.800 & 19$^o$-44$^o$ &1996 Nov 13 & ? & 44  & 23  & - & $<$0.004 & -& $>$114&k,$\pi$\\
V426 Oph & DN ZC & 6.847 & 59$^o\pm$6$^o$ &1994 Sep 18 & Q & 41  & 17  & 0.40 & 0.36 & 2.05&100 &m,$\rho$\\
V345 Pav$\dagger*$ & NL UX & 4.754 & $>$70$^o$& 1994 Sep 23 & ? & 48  & 25  & - & $<$0.10 &-& 630$\pm$100&n,$\sigma$\\
LS Peg & NL DQ? & 4.19 & $\approx$30$^o$ &1998 Nov 23 & ? & 76  & 28  & 0.03 & 0.02& 1.97 &- & 7,$\tau$ \\
RU Peg & DN UG & 8.990 & 33$^o\pm$5$^o$ &1994 Jun 27 & OB & 43  & 18  & 0.08 & 0.06 & 0.50&287$\pm^{23}_{20}$ & d,$\beta,\gamma$\\
KT Per$\dagger*$ & DN ZC & 3.905 & - & 1998 Jan 24 & OB & 9  & 7  & - & $<$0.005 &-& 245$\pm$100 &p\\
CP Pup$\dagger$ & Na SH? & 1.474 & 30$^o\pm$5$^o$& 1998 Nov 06 & ? & 121  & 48  & - & 0.02 &0.90& $>$184 &q,$\upsilon$\\
WZ Sge & DN SU & 1.361 & 75$^o\pm$2$^o$& 1996 May 15 & Q & 82  & 38  & 0.11 & 0.08&0.62&43.5$\pm0.3$ &8,d,$\phi$ \\  
EI UMa & DN UG & 6.434 & - & 1995 Apr 14 & ? & 41  & 23  & 0.52 & 0.33 & 1.17 &-&\\
SU UMa & DN SU & 1.832 & 44$^o\pm$8$^o$&1997 Apr 12 & OB & 47  & 21  & 0.08 & 0.05 & 0.54&260$^{+190}_{-90}$ & b,$\chi$\\
CU Vel$\dagger$ & DN SU & 1.884& 59$^o\pm$6$^o$ & 1994 May 31 & Q & 236  & 94  & - & 0.009 & 0.48 & - & $\psi$\\
IX Vel & NL UX & 4.654 &  60$^o\pm$5$^o$&1995 Oct 29 & HS & 60  & 21  & 0.22 & 0.12 & 0.5& 96$\pm^{10}_{\phantom{0}8}$ & a,$\omega$\\
\hline
\end{tabular}
\end{center}
$\dagger$ serendipitous observation (GIS only)\\
$*$ Non-detections
\end{table*}

\subsubsection{Time series extraction}
\label{sec:lc}
Lightcurves were extracted without energy cuts from all four
detectors for all of our observations. We also extracted
hardness-ratio 
time series from the GIS detectors, using energy ranges of
0.6--2.3\,keV and 2.3--10\,keV. 
We used source extraction circles with radii depending on
brightness of the source in the range 6--19\,arcmin in the SIS 
instruments and 4--11\,arcmin in the GIS instruments. Background rates
were estimated using most of the remainder of the detector. 
Light curves were binned into 16\,s bins for power spectral
analysis. The light curves plotted in this paper have been rebinned
into an integer number of bins per observing interval, with bin sizes
in the range 0.25--4\,ks (depending on count rate).

\subsubsection{Spectral extraction}
\label{sec:spectra}
X-ray spectra were extracted for all four instruments using the same
extraction regions as for the time series. Spectra from the two SIS
instruments were combined, as were spectra from the two GIS instruments. The
resulting spectra were binned to a minimum of 20 counts per bin,  
so that the $\chi^2$ statistic could be used to
test the goodness-of-fit of our models \cite{yaqoobstats}. 
During the later years of the mission the CCD detectors became
gradually degraded, and the low-energy response became increasingly
uncertain. We therefore applied a minimum energy cut to the SIS spectra
that increased from 0.6\,keV to 0.8\,keV during the course of the
mission. We also applied a more extreme cut at 1.2\,keV in the case of
T~Leo, where there is an apparent inconsistency in the calibration of
the SIS and GIS spectra.
Response matrices were generated for each SIS spectrum, since the
spectral resolution of the SIS instruments degraded substantially
during the course of the mission. The standard response matrix was
used for all GIS observations (v1.4). 

\section{The sample}

\subsection{Selection criteria and count rates}
In order to find all \asca\ observations of cataclysmic variables we
cross-correlated the 
\asca\ observations catalogue held by the Leicester Database and
Archive Service (LEDAS) with the Ritter \& Kolb catalogue
\cite{ritter98}. 
Correlations were accepted within a
30\,arcmin search radius. 

\begin{figure}
\centering
\scalebox{0.6}{\includegraphics{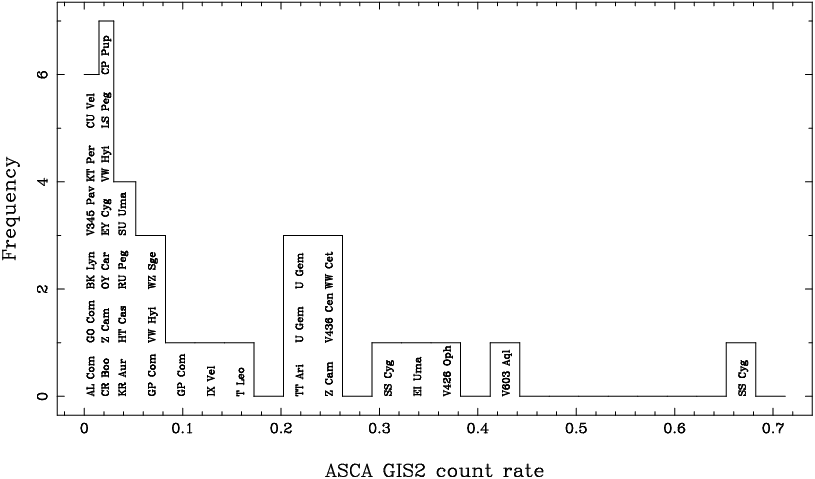}}
\caption{Count rates of the non-magnetic cataclysmic variables observed with the \asca\ GIS2 instrument. Where multiple observations have been made, they have been plotted separately.}
\label{fig:fluxfreqdist}
\end{figure}

In total we found 85 observations of cataclysmic variables. Of these,
46 observations were of magnetic cataclysmic variables 
(17 observations of polars and 29 of intermediate polars) and 5
observations were of super-soft sources. These observations have been
omitted from our sample. This leaves 23 observations of dwarf novae,
of which 4 are repeat observations of a source, and 11 observations of
novae and nova-likes, one of which is a repeat observation. A log of
the \asca\ observations of non-magnetic cataclysmic variables is
presented in 
Table~\ref{table:general} together with system parameters, the optical
state at the time of the observation (if available) 
the mean SIS0 and GIS2 count
rates (without energy cuts, and without corrections for vignetting), and the GIS2 hardness ratio (the ratio of counts above and below 2.3\,keV). 
We note that 25 out of the 29 non-magnetic cataclysmic variables 
were detected, including four serendipitous GIS-only
detections. 
The mean GIS2 count rates are plotted as a 
histogram in Fig.~\ref{fig:fluxfreqdist}. It can be seen that SS~Cyg has the
highest observed flux of any of the
non-magnetic cataclysmic variables, even though it was observed during
outburst when the X-ray count rate is expected to be suppressed by a
factor of $\sim$5 (\ncite{Wheatley03}). The fainter observation of SS~Cyg was made in an
exceptionally faint quiescent state (see Sect.\,\ref{sec:sscygtemporal}).
For each observation, the optical state was determined through inspection of AAVSO
visual light curves (where available) 
and these are summarised in Table\,\ref{table:observedstates}.

\begin{table}
\caption{Summary of the optical states of non-magnetic cataclysmic variables
observed with ASCA. }
\label{table:observedstates}
  \centering
    \begin{tabular}{ccrr}
         &  & No.\  of & No.\  of \\
Type     & Optical state &  targets & obs.\\
\hline
Dwarf novae         & all states & 19 & 23 \\  
                    & outburst   &  5 &  6 \\  
                    & quiescence & 12 & 14 \\  
                    & unknown    &  3 &  3 \\
Novae \& nova-likes & all states & 10 & 11 \\
                    & high-state &  5 &  5 \\
                    & unknown    &  5 &  6 \\
\end{tabular}
\end{table}

\subsection{Hardness ratios} 
\label{sec:ratio}

\begin{figure}
\centering
\scalebox{0.6}{\includegraphics{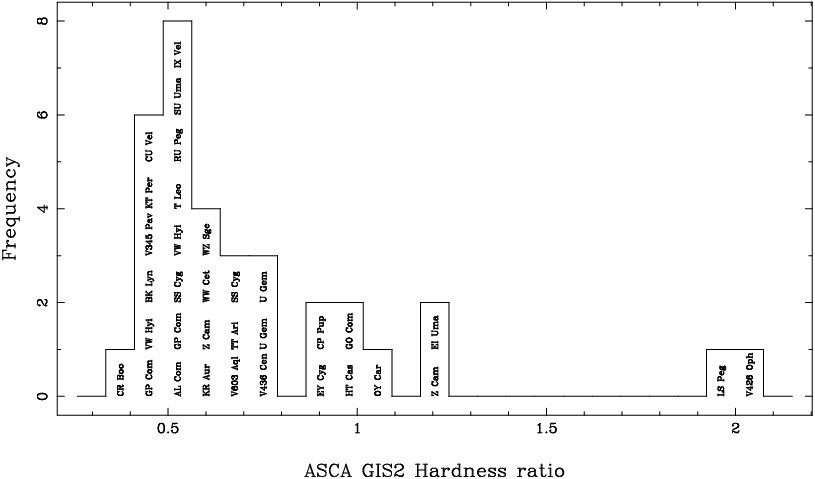}}
\caption{Histogram of the \asca\  GIS2 hardness ratio for all our
observations. }
\label{fig:hardnessfreqdist}
\end{figure}

\begin{figure}
\centering
\includegraphics[width=8.4cm]{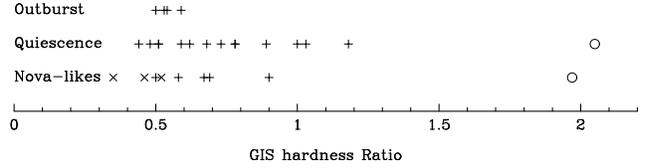}
\caption{Plot of the GIS2 hardness ratios by system class and outburst
state (if known). The crosses indicate AM CVn systems.  The two systems with extreme hardness in our sample, V426 Oph and  LS~Peg, are indicated with circles.}
\label{fig:ratiostates}
\end{figure}

Figure\,\ref{fig:hardnessfreqdist} shows the distribution of 
\asca\  GIS hardness ratios. 
The systems LS~Peg and V426~Oph stand out with exceptionally hard
X-ray spectra. We discuss these systems in Sect.\,\ref{sec:ips} and
argue that they should be re-classified as intermediate polars. 
The hardness ratios are also plotted by type and optical
state in Fig.\,\ref{fig:ratiostates}. It can be seen that all dwarf novae
observed in outburst have relatively soft X-ray spectra. 
Our spectral analysis of Sect.\,\ref{sec:spectral} shows that this is
due to an increased contribution by cool gas.  
Dwarf novae in quiescence exhibit a much wider range of hardness.  Nova-likes appear to be intermediate between the two.
  In order to make quantitative comparisons between the distributions, we have carried out Kolmogorov-Smirnov tests (excluding LS~Peg and V426~Oph).  This shows that the nova-like hardness distribution is most similar to the outburst hardness distribution, with a KS probability of 0.79.  There is less similarity between the outburst and quiescent hardness distributions (KS probability of 0.25), or the quiescent and nova-like hardness distributions (KS probability of 0.45).

\subsection{X-ray identification of V345~Pavonis}

V345~Pav (EC\,19314-5915) is an optically  
bright eclipsing cataclysmic variable
discovered in the Edinburgh-Cape blue object survey \cite{v345pav}. 
Buckley et al.\   suggest that V345~Pav is the optical counterpart of
the \heao\   X-ray source 1H1930-589.  However, the \asca\  GIS image shows
that the \heao\   source actually lies 9\,arcmin from the position of V345 Pav. 
 The brighter X-ray source is coincident
with the bright star CD-59 7229 (V=10.5). This star has a B-V colour
of +1.1 indicating that it may be a late-type coronally-active star.

\section{Timing Analysis}
We analysed the temporal behaviour of our sample of \asca\ observations
beginning at the longest available timescales and moving progressively
to shorter timescales. The longest available timescales are years, due
to repeat observations of individual systems. The shortest timescales are
set by the time-resolution of our individual observations, usually 16\,s.

\subsection{Repeat observations}
Four of the systems in our sample were observed more than
once with \asca , allowing us to detect variations over long
timescales. 

Z~Cam was observed with \asca\  both in outburst and during a
transition to outburst. During the transition the \asca\  count rate 
dropped rapidly as the X-rays were suppressed. During outburst the \asca\
count rate remained low. These observations have been discussed in
detail by \scite{baskill01}.  U~Gem was observed twice in quiescence,
with a similar count rate in both cases. These observations were
discussed by \scite{ugemszkody}.
The repeat observations of SS~Cyg and VW~Hyi have not been presented
elsewhere, and so we describe them here in more detail. 

\subsubsection{SS Cygni}
\label{sec:sscygtemporal}

\begin{figure*}
\centering
\scalebox{1.1}{\includegraphics{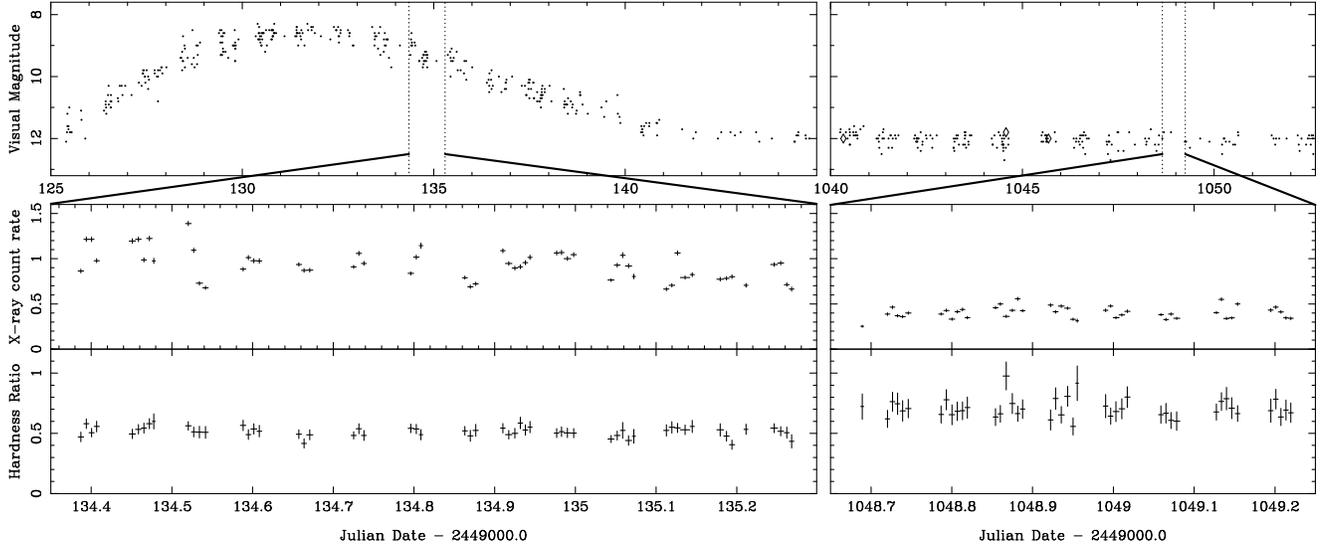}}
\caption{Optical lightcurve (top), X-ray lightcurve (middle) and
X-ray hardness ratio (bottom) of SS~Cyg during the two \asca\
observations.  The X-ray count rate is found to be unusually low in
quiescence. 
}
\label{fig:sscygaavsoxray}
\end{figure*}

SS~Cyg was observed twice with \asca,  with two and a half years
between the observations.  It was the only dwarf nova
to be observed in both outburst and quiescence with \asca.  
The outburst data have been published previously by \scite{ascasscyg}
and \scite{do97} but the quiescent observation has not been presented 
elsewhere. 

The \asca\ lightcurves of SS~Cyg are shown in Fig.\,\ref{fig:sscygaavsoxray}.
Remarkably the count rate was higher in outburst than
in quiescence. This is in contrast to all previous X-ray observations
of SS~Cyg \egcite{Ricketts79,Jones92,Wheatley03}. 

Comparison of our \asca\  fluxes with previous observations of
SS~Cyg (assuming our fitted spectra of Sect.\,\ref{sec:spectral}) shows that
the outburst brightness was normal, and that the quiescent brightness was
unusually faint. For instance, \scite{Wheatley03} found 3--20\,keV
fluxes of 1--4$\times10^{-11}\rm\,erg\,s^{-1}\,cm^{-2}$ in outburst
and 10--16$\times10^{-11}\rm\,erg\,s^{-1}\,cm^{-2}$ in quiescence. 
Integrating our best fitting spectral models over this range yields
\asca\ fluxes of 3$\times10^{-11}\rm\,erg\,s^{-1}\,cm^{-2}$ and
2$\times10^{-11}\rm\,erg\,s^{-1}\,cm^{-2}$ respectively. 
None of the previous X-ray observations of SS~Cyg show such a low flux in quiescence. 

Figure\,\ref{fig:sscygaavsoxray} also shows the hardness ratio of SS~Cyg
during these two observations. It can be seen that the hardness ratio
was constant in both observations, but that the outburst observation was
softer than the quiescent observation. This behaviour is
consistent with that seen in previous observations of this object 
\egcite{Wheatley03} and of other non-magnetic cataclysmic variables
(Fig.\,\ref{fig:ratiostates}). 

\subsubsection{VW Hydri}

\begin{figure*}
\centering
\scalebox{1.1}{\includegraphics{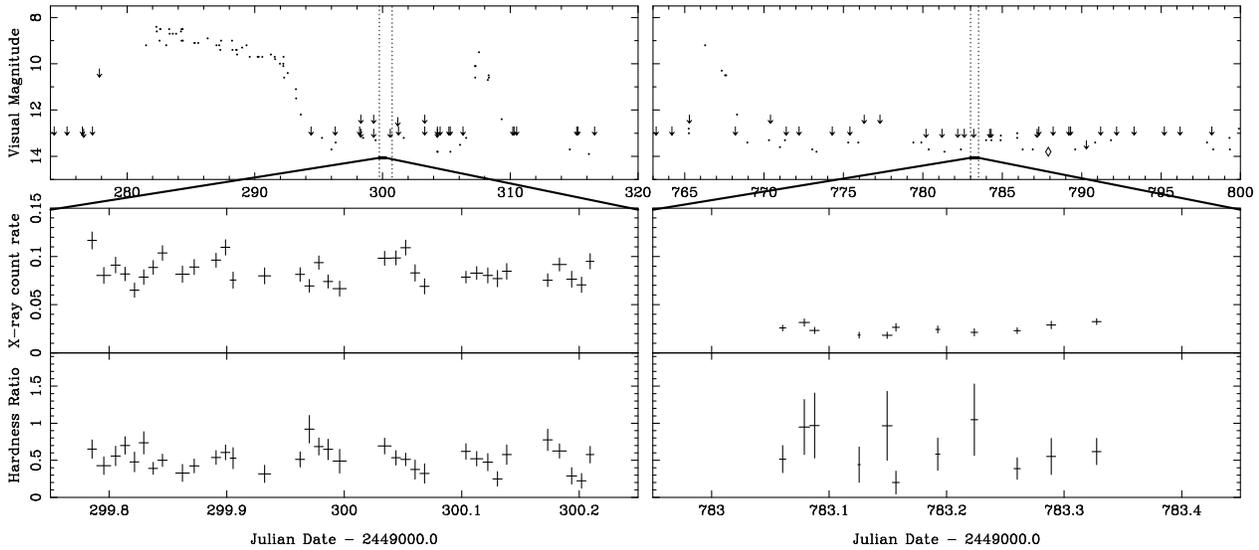}}
\caption{
Optical lightcurve (top), X-ray lightcurve (middle) and
X-ray hardness ratio (bottom) of VW~Hyi during the two \asca\
observations.  The X-ray count rate differs by almost an order of
magnitude, even though both observations were made during quiescent
intervals. 
}
\label{fig:vwhyiaavsoxray}
\end{figure*}

\begin{table}
\caption{Comparison of X-ray fluxes of VW~Hyi in quiescence. All fluxes are in units of $10^{-12}\,\rm erg\,s^{-1}\,cm^{-2}$. }
\label{table:vwflux}
  \centering
    \begin{tabular}{lcrrr}
      & Bandpass &  & \multicolumn{2}{c}{ASCA Fluxes}  \\
Inst. & [keV] & Flux & Obs.\,1 & Obs.\,2 \\
\hline
EXOSAT      &  0.04-\phantom{1}6& 15$^a$ & 6.1 & 1.6 \\
ROSAT+Ginga & 0.04-10& 21$^b$ & 6.8 & 1.9 \\
BeppoSAX    &  0.1\phantom{4}-10& 15$^c$ & 6.5 & 1.8 \\
XMM-Newton  &  0.2\phantom{4}-12& 6$^d$& 6.2 & 1.8 \\
\end{tabular}\\
$^a$ \protect\scite{vwhyiexosat}
$^b$ \protect\scite{wheatley96a}\\
$^c$ \protect\scite{vwhyibepposax}
$^d$ \protect\scite{pandel03}
\end{table}

VW~Hyi was observed twice with \asca , 
but neither observation has been published. 
We present both \asca\  lightcurves of VW~Hyi in 
Fig.\,\ref{fig:vwhyiaavsoxray}.  
Both observations were made during quiescence, but there is a
four-fold difference in the X-ray count rates. This behaviour
demonstrates the value of X-ray observations of such systems. While
the optical emission is sensitive to the global state of the
accretion disc, it is clear that there must also be activity in the inner
disc that does not correlate with the optical emission. 
The hardness ratios in the two observations are consistent ($0.52\pm0.12$ in the earlier observation, $0.66\pm0.31$ in the later), showing
that the difference between the two observations was probably due to a
changing accretion rate.

In Table\,\ref{table:vwflux} we compare our observed ASCA fluxes with 
previous X-ray observations. In each case we calculate ASCA fluxes 
in the same energy range as the published fluxes using our best-fitting 
spectral model of Sect.\,\ref{sec:spectral}.  VW~Hyi was 
fainter during our first ASCA observation than during most previous 
observations, but the measured flux is consistent with that seen by 
\scite{pandel03} with XMM-Newton. The flux during our second ASCA observation 
is much fainter than any of the previous X-ray observations of VW~Hyi.

Figure\,\ref{fig:vwhyiaavsoxray} shows that the bright ASCA
observation was made in a short period of quiescence between a
superoutburst and a normal outburst, whereas the faint observation was
made in the middle of a longer quiescent interval. One might imagine that 
the difference between these observations is related to their
relative proximity to outbursts. 
However, both the XMM-Newton and ROSAT/Ginga observations 
were also made in the middle of quiescent intervals. It is clear that there 
is no simple relationship between X-ray flux and inter-outburst phase. 

\subsection{Long-timescale variations within observations}
\label{sec:longperiodtrends}
In order to search for overall brightness variations on the timescale of our
observations we carried out a least-squares linear fit to all our
SIS0 lightcurves. Our sensitivity is not the same in each case, due to
the range of observation duration and source brightness, but we do
detect overall brightness variations in four cases. 

\begin{figure}
\centering
\includegraphics{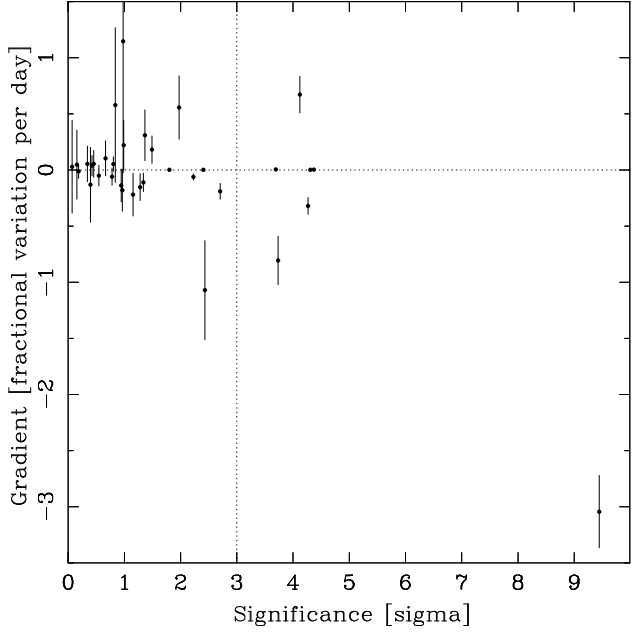}
\caption{Best fit gradients for all our SIS0 lightcurves versus the
significance of their deviation from constancy (in
$\sigma$). Gradients are normalised by count rate, so represent the
fractional variation per day.}
\label{fig:grad}
\end{figure}

The best fit gradient for each observation is plotted in 
Fig.\,\ref{fig:grad}. Seven observations have well constrained 
gradients at the 3-$\sigma$ level. However, three of these exhibit 
fractional variations
that are
less than one per cent per day, and we regard these as
essentially constant. The remaining four systems all have gradients
greater than thirty per cent per day, and these systems are listed in 
Table\,\ref{tab:trends}.
All four are dwarf novae, three of which were in outburst during the
observation. With three of the the six outburst observations showing
an overall trend, and only one of the fourteen quiescent observations,
it is clear that the outburst state is most closely associated with 
large scale X-ray flux variations. 

\begin{table}
\caption{Observations for which a gradient was detected in the SIS0
lightcurve.}
\label{tab:trends}
\begin{center}
\begin{tabular}{lccccc}
Target & State & Duration & Counts$^a$  & Gradient$^b$ & Sig . \\
\hline
Z Cam  & T  & 36\,ks & 0.39\,s$^{-1}$ & $-3.0\pm$0.3 & $9.4\sigma$ \\
SS Cyg & OB & 78\,ks & 1.46\,s$^{-1}$ & $-0.3\pm$0.1 & $4.3\sigma$ \\
SU UMa & OB & 47\,ks & 0.08\,s$^{-1}$ &  $0.7\pm$0.2 & $4.1\sigma$ \\
WW~Cet & Q  & 41\,ks & 0.44\,s$^{-1}$ & $-0.8\pm$0.2 & $3.7\sigma$ \\
\end{tabular}
\end{center}
$^a$ SIS0 count rate.\\
$^b$ Fractional variation per day.\\
\end{table}

\begin{figure}
\centering
\scalebox{1.1}{\includegraphics{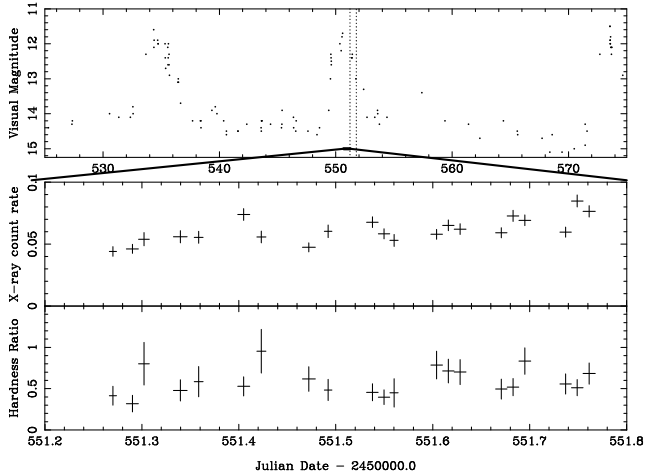}}
\caption{Optical lightcurve (top), X-ray lightcurve (middle) and
X-ray hardness ratio of SU~UMa during the \asca\ observation.}
\label{fig:suumaaavsoxray}
\end{figure}

\begin{figure}
\centering
\scalebox{1.1}{\includegraphics{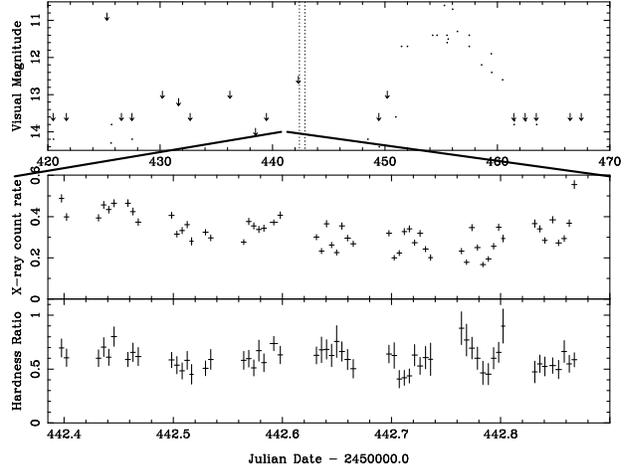}}
\caption{Optical lightcurve (top), X-ray lightcurve (middle) and
X-ray hardness ratio of WW Cet during the \asca\ observation.
This is the only case in which an overall flux variation was detected
in a dwarf nova in quiescence. 
}
\label{fig:wwcetaavsoxray}
\end{figure}

The second observation of Z~Cam was made during a transition to outburst and showed a
dramatic decrease in the \asca\ count rate (see also \ncite{baskill01}). 
This is normal behaviour for dwarf novae.
The observation of SS~Cyg, made towards the end of an outburst,
also shows a decreasing count rate (Fig.\,\ref{fig:sscygaavsoxray}), 
although in this case the variation does
not appear to be part of the outburst to quiescence transition.
SU~UMa was observed to be brightening during the second half of
the outburst, and its lightcurve is plotted in Fig.\,\ref{fig:suumaaavsoxray}.
The only overall trend detected in a quiescent dwarf nova 
was in WW~Cet, which faded during its 13\,h observation.  
The \asca\  lightcurve of WW~Cet is shown in Fig.\,\ref{fig:wwcetaavsoxray}.
The observation was made eight days before an outburst. WW~Cet has a well defined outburst recurrence time of around sixty days. 

\begin{figure}
\centering
\includegraphics{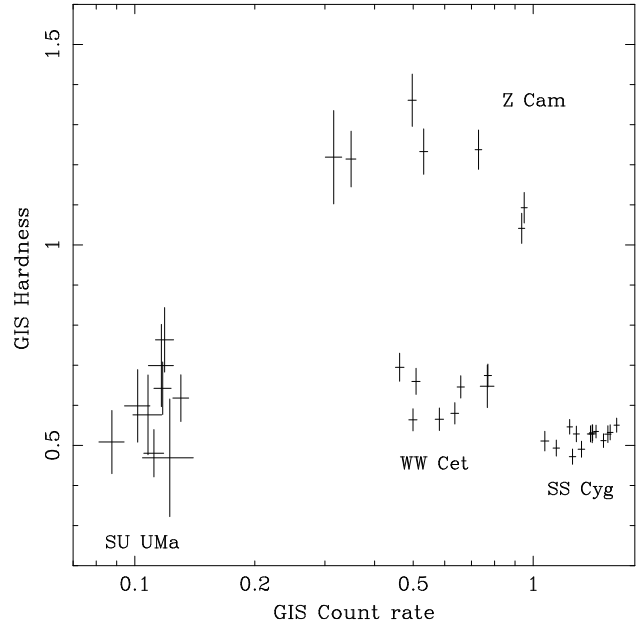}
\caption{\asca\ GIS hardness as a function of count rate for the four
observations in which overall flux variations are observed. }
\label{fig:cr_hr}
\end{figure}

Figure\,\ref{fig:cr_hr} shows the GIS hardness ratio versus count
rates for all four systems. 
In Z~Cam the hardness ratio increased during the early part
of the X-ray suppression, and in SS~Cyg the hardness ratio dropped
slightly during the decline, but in general the spectra remained stable
during these flux variations. 

\subsection{Orbital modulation}
\label{sec:orbitalmodulations}
We have calculated power spectra for all our \asca\  lightcurves of
non-magnetic cataclysmic variables. We used the Lomb-Scargle
algorithm, as implemented by \scite{powerspec}, but with a slight
modification to the normalisation (such that the power spectrum is
normalised by expected variance rather than measured variance, which
can be dominated by signal). 
Some of these power spectra are dominated by red noise at low
frequencies, and this limited our ability to search for low frequency periodic
modulations (see Sect.\,\ref{sec:shortperiodmodulations}). 
However, since we know {\it a priori} the orbital period of each of
our systems, we circumvented the problem of red noise by searching
our power spectra for peaks centred precisely on the known orbital periods. 
Of course, this approach does not entirely remove the possibility of
contamination by red noise. 

Inspecting the power spectra of our thirty light curves we found 
peaks centred on the orbital period in
seven cases. These were observations of OY~Car (27.5 cycles covered), 
HT~Cas (12.9\,cycles), V436~Cen (12.2\,cycles),
T~Leo (14.2\,cycles), V426~Oph (1.7\,cycles), EI~UMa (1.8\,cycles) and
both observations of U~Gem (4.6 \& 3.9\,cycles). All are dwarf
novae observed in the quiescent state, with the exception of
EI~UMa for which optical observations were  not available. 
The observations of V426~Oph and EI~UMa were not sufficiently long to
be sure that these modulations were periodic, but these observations
nevertheless do exhibit power close to the known orbital period.
The power spectrum of V426~Oph is discussed further in 
Sect.\,\ref{sec:v426periods}.

It is striking that this set includes three of the four systems with
the highest known inclination angles: OY~Car, HT~Cas and U~Gem, with
inclinations of 83.3$\pm$0.2, 81$\pm$1 and 69.7$\pm$0.7\degrees
respectively \cite{oycar89,htcasinc,ugeminc}. 
WZ~Sge is the only system in our sample known to have a
correspondingly high inclination but which did not exhibit an orbital
modulation.  Of the remaining systems
V436~Cen and V426~Oph may also have relatively high inclinations of 
65$\pm$5\degrees and 59$\pm$6\degrees respectively
\cite{ritter98,v426oph88}. 
None of the systems in this \asca\ sample with suspected orbital modulation are known to have an inclination angle less than 60\degrees. 


\begin{figure*}
\centering
\includegraphics{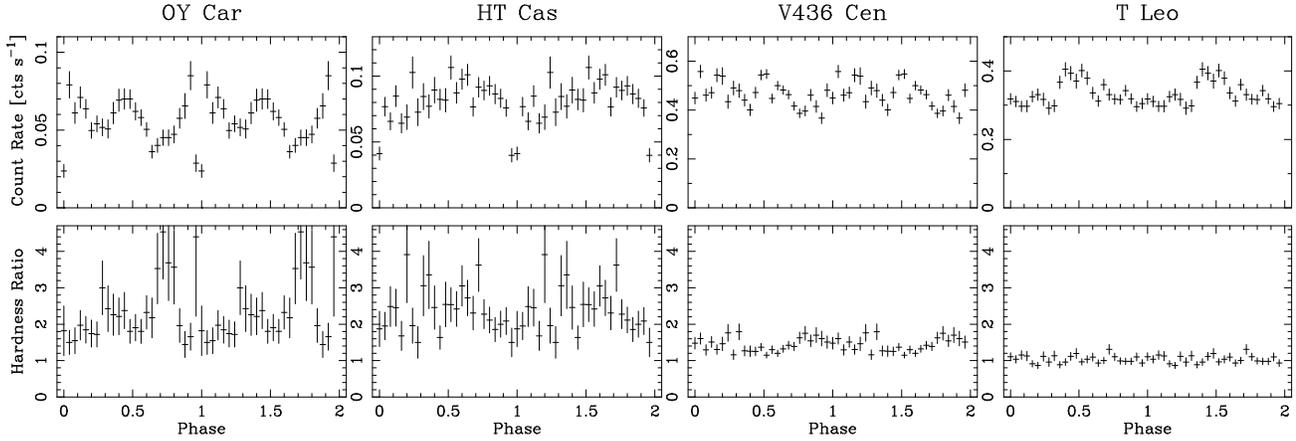}
\caption{The \asca\ SIS0 lightcurves and hardness ratios of
four of the seven systems for which we find evidence of an orbital
modulation. From left to right: OY~Car, HT~Cas,
V436~Cen, T~Leo. }
\label{fig:orbital}
\end{figure*}

\begin{figure}
\centering
\includegraphics{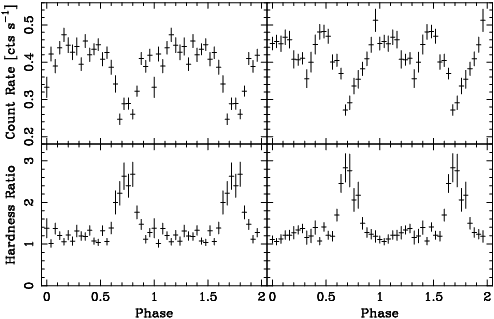}
\includegraphics{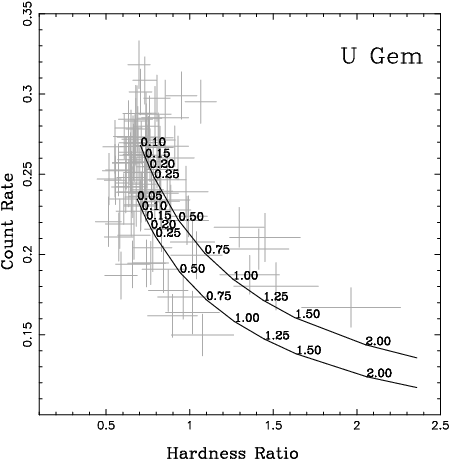}
\caption{Top: the U~Gem lightcurves and hardness ratios from the two
observations folded on the orbital period. 
Bottom: count rates plotted against hardness for both 
observations. The solid curves represent the relationships expected
for the best fitting spectral models of Sect.\,\ref{sec:complex} and a 
range of photo-electric absorption cross-sections (labels are in units of 
10$^{22}$cm$^{-2}$).}
\label{fig:ugemvariations}
\end{figure}


The folded GIS1+2 lightcurves and hardness ratios for OY~Car, HT~Cas,
V436~Cen and T~Leo are presented in Fig.\,\ref{fig:orbital}.
OY~Car and HT~Cas are eclipsing systems, and X-ray eclipses are
apparent in both folded lightcurves (see also
\ncite{htcas,oycareph,oycarxmmpj}). Both systems also show
orbital modulation in addition to the eclipses. In the case of OY~Car
this is anti-correlated with the hardness ratio, suggesting an origin
in photo-electric absorption. A similar anti-correlation is also found
in lightcurves and hardness ratios of the two observations of U~Gem 
(Fig.\,\ref{fig:ugemvariations}, and \ncite{ugemszkody}). 
In Fig.\,\ref{fig:ugemvariations} we also plot
count rate against hardness ratio and overlay lines of varying absorption
column density for the best fit emission spectrum of
Sect.\,\ref{sec:complex}. It can be seen that the variations are
broadly consistent with photo-electric absorption. 
We note that the orbital absorption dips were originally discovered in 
U~Gem by \scite{mason88} during an outburst. The ASCA observations are the first in which the absorption dips were detected in quiescence 
\cite{ugemszkody}. 
\scite{Mauche03} have modeled \chandra\ grating spectra of the dips in U~Gem 
and found an acceptable fit with a partial-covering absorber. 

In other systems we find that the orbital modulation was not
anti-correlated with hardness. The best example is T~Leo
(Fig.\,\ref{fig:orbital} and \ncite{tleoasca}) 
but the anti-correlation with hardness is also
missing in HT~Cas and V436~Cen. In V426~Oph and EI~UMa the orbital
phase coverage was too poor to draw firm conclusions. 
Orbital X-ray modulation has not been previously claimed in V436~Cen, 
V426~Oph, or EI~UMa.

\subsection{Short timescale modulations}
\label{sec:shortperiodmodulations}
The discovery of periodic X-ray modulation (other than the orbital period) 
is the defining characteristic of intermediate polars. We therefore do
not expect to find such periodicities in our sample of non-magnetic
cataclysmic variables. 
Nevertheless, 
our large set of relatively high signal-to-noise lightcurves may 
be expected to contain features previously missed with less sensitive
instruments. 
We are encouraged by the detection of a 37\,min period in OY~Car with XMM-Newton 
\cite{oycarxmm2}.

The search for periodic modulations in our power spectra was
complicated by the presence of red noise that often dominates the
low frequencies. At higher frequencies the power spectra are dominated
by white noise, and we could apply the standard tests for significance
of peaks in the power spectrum \cite{powerspec}. 

An inspection of the power spectra of our thirty light curves revealed
four candidate periodicities. 
In each case we required that the peak satisfied the
significance test used by \scite{powerspeceinstein} for \einstein\ data, and also that it be well separated from obvious red noise peaks. 
We found candidate periodicities in the power spectra of 
WW~Cet, V426~Oph, LS~Peg, 
and EI~UMa.  
The SIS0 power spectra are plotted in 
Figs.\,\ref{fig:wwcetperiods},\,\ref{fig:v426periods},\,\ref{fig:lspegperiods}\,\&\,\ref{fig:eiumaperiods}, and the folded SIS0 lightcurves are presented in Fig.\,\ref{fig:periods}.

\subsubsection{WW~Cet} 
\begin{figure*}
\centering
\includegraphics{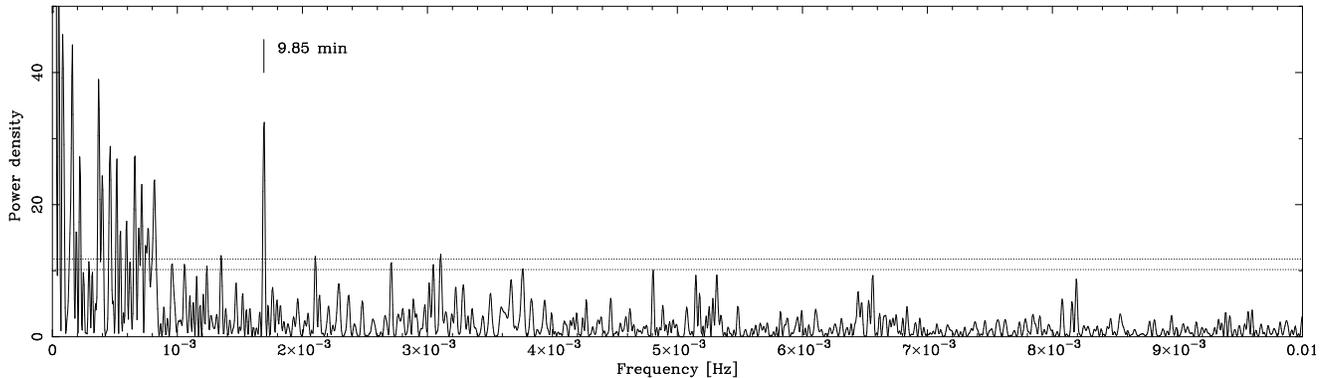}
\caption{SIS0 power spectrum of the \asca\  observation of WW~Cet. 
The horizontal dotted lines indicate the 95 and 99 per confidence
limits for detection of significant power above white noise. The labelled peak
may represent a true periodicity, but it could instead be a feature of
the red noise spectrum.} 
\label{fig:wwcetperiods}
\end{figure*}

Significant power was detected at a period
of 9.85$\pm0.05$\,min in the SIS0 power spectrum of WW~Cet 
(see Fig.\,\ref{fig:wwcetperiods}). Fitting a sinesoidal function to the
folded SIS0 lightcurve we found an amplitude of 11$\pm$1 per cent (where
amplitude is the half amplitude of the sine curve divided by the mean). 
The folded SIS0 lightcurve and best-fit sine function are presented in 
Fig.\,\ref{fig:periods}.

The same peak was seen in the power spectrum of the
GIS2 lightcurve, and weakly in the power spectrum of the SIS1
lightcurve, but it was not seen in the GIS3 lightcurve. 
This difference in the four instruments is due to vignetting resulting from
misalignment of the \asca\ telescopes. 
For targeted observations the count rates were highest
in the SIS0 and GIS2 instruments and lowest in the SIS1 and GIS3
instruments. Folding on the SIS0 period we found 
the GIS3 lightcurve to be consistent with our sine fit to the SIS0
lightcurve. 
We note, however, that the power spectrum of WW~Cet 
contains red noise that dominates at frequencies below
0.001\,Hz. 
Without a thorough understanding of the form of this noise
it is difficult to assess whether the detected peak represents a truly period
modulation.

\subsubsection{V426 Oph}
\label{sec:v426periods}
\begin{figure}
\centering
\scalebox{0.5}{\includegraphics{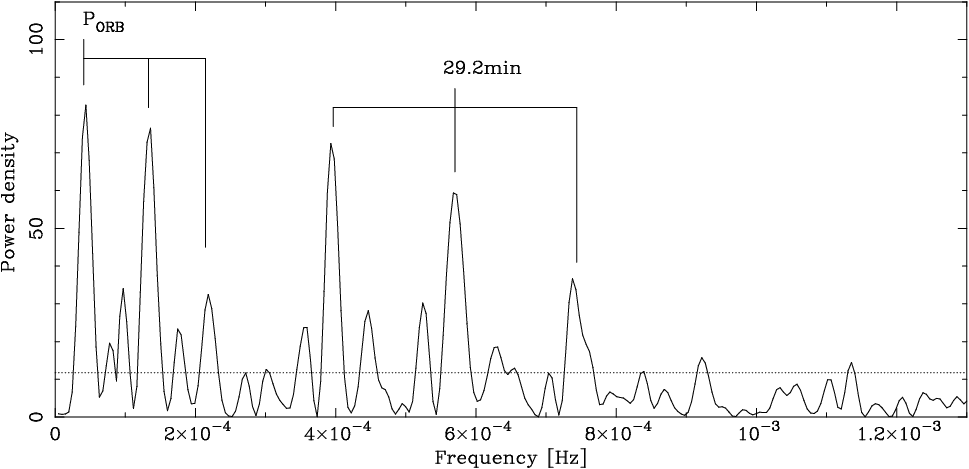}}
\caption{SIS0 power spectrum of the \asca\  observation of V426~Oph. 
The horizontal dotted line indicates the 99 per confidence
limits for detection of significant power above white noise. Proposed real periodic
modulations are labelled, together with probable sampling aliases
separated by the \asca\ orbital frequency.}
\label{fig:v426periods}
\end{figure}

At first sight the power spectrum of V426~Oph looks like it is
dominated by red noise (Fig.\,\ref{fig:v426periods}). However, closer
inspection reveals that several of the peaks are separated by
precisely the orbital frequency of the \asca\ satellite.
Since our lightcurves have gaps on the \asca\ orbital period (96\,min) 
we can expect to see power at beat frequencies between real periods
and the sampling period. In the power spectrum of 
Fig.\,\ref{fig:v426periods} it can be seen that there is power at the
orbital frequency of V426~Oph, and that two of the other low frequency peaks
can be associated with the beats between this orbital frequency and
the \asca\ sampling frequency. Also, if we assume the peak at 
29.2$\pm$0.9\,min represents a real modulation then the two peaks 
either side of it can also be interpreted as sampling aliases.

We conclude that the X-ray emission of V426~Oph is 
modulated on a period of 29.2$\pm$0.9\,min, with excess power suggestive of orbital modulation. 
Fitting the SIS0 lightcurve folded at 29.2\,min
yielded an amplitude of 20$\pm$1 per cent. 
The folded SIS0 lightcurve and best-fit sine function 
are presented in Fig.\,\ref{fig:periods}.
In Sect.\,\ref{sec:ips} we argue that V426~Oph should be
reclassified as an intermediate polar based independently on its exceptionally 
hard X-ray spectrum.

V426~Oph has previously been claimed to be an intermediate polar 
by \scite{v426ophip} but this evidence was later refuted by
\scite{v426ophip?}. \scite{v426ophginga} found evidence for a 28\,min
modulation in the \ginga\ lightcurve, and it seems likely that this is
the same modulation as detected in the \asca\ lightcurve. We note that 
\scite{v426ophginga} believed the 28\,min modulation was not strictly
periodic.

\subsubsection{LS~Peg}
\begin{figure*}
\centering
\includegraphics{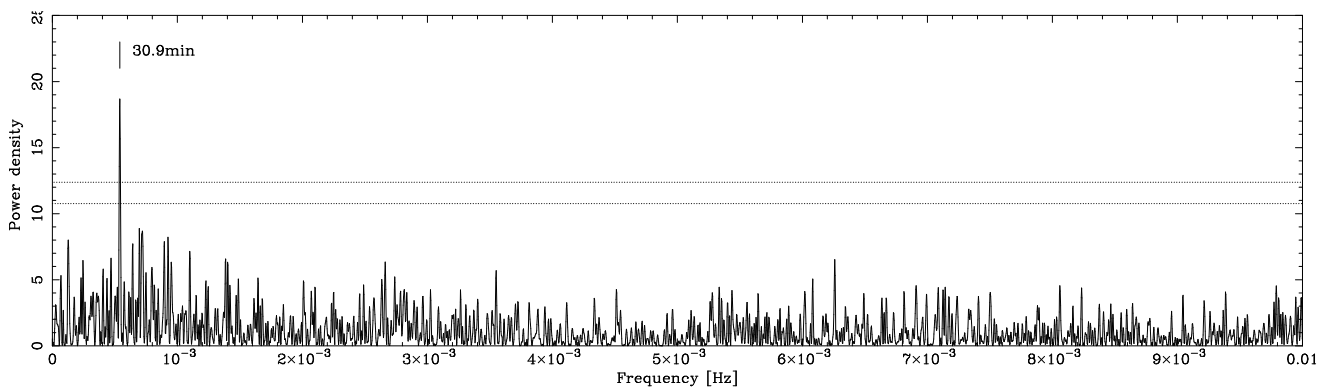}
\caption{SIS0 power spectrum of the \asca\  observation of LS~Peg. The
horizontal dotted lines indicate the 95 and 99 per cent confidence
limits for the detection of significant power above white noise.}
\label{fig:lspegperiods}
\end{figure*}

Significant power at a period of 30.9$\pm$0.3\,min was detected in the
SIS0, GIS2 and GIS3 power spectra of LS~Peg 
(see Fig.\,\ref{fig:lspegperiods}). 
The power spectra appear to be relatively free from red noise. 
Fitting the folded SIS0 lightcurve with a sine function yielded an
amplitude of 32$\pm$5 per cent. The folded SIS0 lightcurve  and
best-fit sine function are presented in
Fig.\,\ref{fig:periods}. The folded SIS1 lightcurve is consistent
with this amplitude. 

\scite{lspegip}  report a detection of a period at 
29.6$\pm$1.8\,min in the circular polarisation of
LS~Peg. The coincidence of these detected periods 
leads us to believe that the modulation detected in the \asca\
lightcurve is truly periodic, and that the modulation in X-rays and 
circular polarisation have a common physical origin. 

In Sect.\,\ref{sec:ips} we argue that
the X-ray spectrum of LS~Peg is much more like that of an intermediate
polar than a non-magnetic cataclysmic variable. 
The detection of periodic modulation also supports the
reclassification of LS~Peg as an intermediate
polar.

\subsubsection{EI~UMa}
\begin{figure*}
\centering
\includegraphics{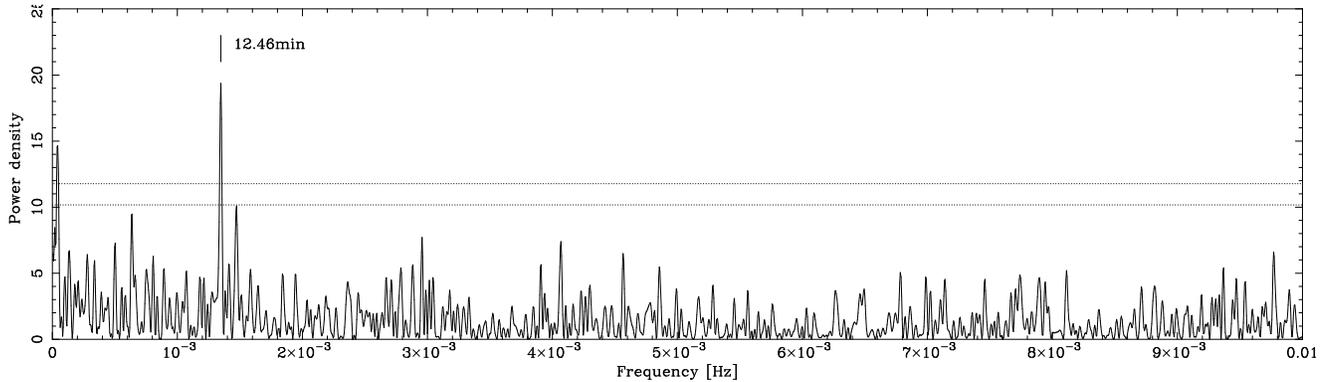}
\caption{SIS0 power spectrum of the \asca\  observation of EI~UMa. The
horizontal dotted lines indicate the 95 and 99 per cent confidence
limits for the detection of significant power above white noise.}
\label{fig:eiumaperiods}
\end{figure*}

\begin{figure*}
\centering
\scalebox{0.5}{\includegraphics{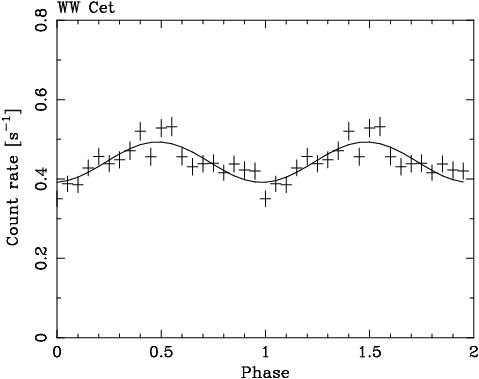}} \hfil
\scalebox{0.5}{\includegraphics{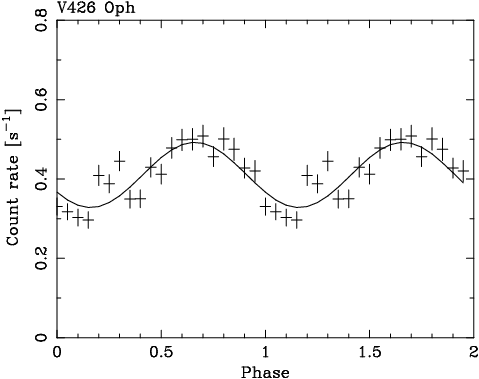}} \hfil
\scalebox{0.5}{\includegraphics{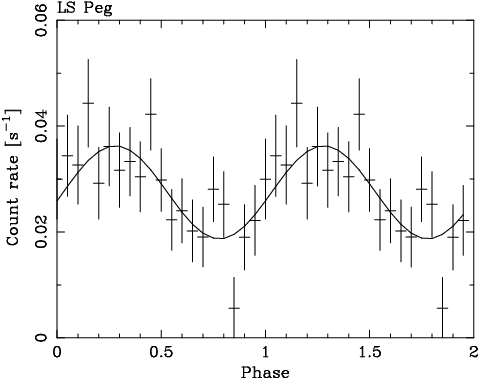}} \hfil
\scalebox{0.5}{\includegraphics{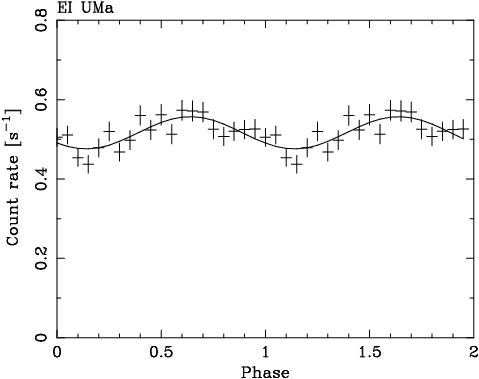}} \\
\caption{The \asca\ SIS0 folded lightcurves of the 
four candidate periods from Sect.\,\ref{sec:shortperiodmodulations}.
The solid curves represent the sine function fits used to measure
pulsation amplitudes. 
From left to right: WW~Cet, V426~Oph, LS~Peg and EI~UMa. 
In the case of WW~Cet, the observed modulation may merely be a
feature of the red noise spectrum. 
}
\label{fig:periods}
\end{figure*}

Significant power was detected in the SIS0 power spectrum of EI~UMa
at a period of 12.36$\pm$0.09\,min (see
Fig.\,\ref{fig:eiumaperiods}). Fitting a sine function to the folded
SIS0 lightcurve we found an amplitude of 8$\pm$1 per cent. 
The folded SIS0 lightcurve and best-fit sine function 
are presented in Fig.\,\ref{fig:periods}.

This period was seen only in the SIS0 lightcurve and not in the other
three instruments. This is probably because the SIS0 count rate is a factor
1.3--3.3 higher than the other instruments. The folded lightcurves of
all three are consistent with our fit to the folded SIS0 lightcurve.

This is the first detection of a non-orbital period from this system.
In Sect.\,\ref{sec:ips}
we argue that our fits to the \asca\  spectra of EI~UMa require it to
be classified, with V426~Oph and LS~Peg, as an intermediate polar. The
discovery of this period provides supporting evidence for this
interpretation. 

\subsection{\asca\ power spectrum of V603~Aql}
The \asca\  lightcurve of V603~Aql is of high quality and is of
particular interest because evidence of periodic modulations have been 
found in a previous X-ray observation \cite{Udalski89}. 
V603~Aql is also notable for 
exhibiting both positive and negative superhumps in its optical lightcurves
\cite{Patterson97}.

\scite{Udalski89} claim the detection of a 63\,min period in the
\einstein\ lightcurve of V603~Aql.  \scite{powerspeceinstein} analyse
the same data and agree that there may be a periodic modulation at
63\,min or the first harmonic of that period at 31$\pm$2\,min, but
conclude that further observations are required to establish the
stability of this possible modulation. 
Modulations at similar periods may also be present in the optical
lightcurves of V603~Aql \cite{Udalski89,Patterson97}. 
However, a recent study using  \rosat\
data found no evidence for an X-ray periodicity \cite{Borczyk03}. 

\begin{figure*}
\centering
\includegraphics{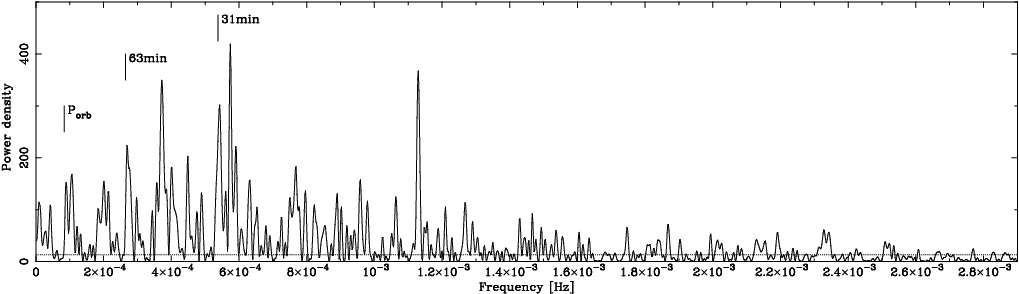}
\caption{SIS0 power spectrum of the \asca\  observation of
V603~Aql plotted in the range 0--0.0029\,Hz. The orbital period is
indicated, as well as two periods claimed from previous X-ray
data. The horizontal dotted line indicates the 99 per cent confidence
limit for detection of significant power assuming a white noise dominated power spectrum. We believe that all the peaks are
probably the product of red noise and do not represent the detection of truly
periodic modulation.}
\label{fig:v603aqlperiods}
\end{figure*}

We plot the \asca\  SIS0 power spectrum of V603~Aql in 
Fig.\,\ref{fig:v603aqlperiods}. It can be seen that there are
prominent peaks close to the orbital period and also close to the
31\,min and 63\,min periods previously claimed in X-rays. 
There are, however, several other prominent peaks, and it is clear
that there is significant power over a wide range of frequencies. We
conclude that the \asca\ lightcurves of V603~Aql are dominated by
intense red noise, and that there is no evidence for periodic
modulation of the X-ray flux. 

\section{Spectral analysis}
\label{sec:spectral}
\asca\ was designed primarily for spectroscopy and was the first
X-ray observatory to be equipped with CCD detectors. Our large sample
of \asca\  spectra of non-magnetic cataclysmic variables represents an
excellent opportunity to study the properties of their X-ray emission
as a whole. \asca\ spectra cover the range 0.6--10\,keV so we were
sensitive to the dominant hard X-ray component of dwarf novae in
quiescence, and the residual hard X-ray emission of dwarf novae in
outburst, but we were not sensitive to the extreme-ultraviolet
component that dominates the luminosity during outburst. 

Our approach was to begin by fitting a very simple model to all our
spectra, and then gradually increase the complexity of the model until
all spectra were well fit. We did not attempt to fit more complex models
to spectra with less than 2000 counts (with the exception of LS~Peg).

\subsection{Spectral fitting methods}
\label{sec:spectralfitting}
We carried out our spectral fitting using the {\it XSPEC} software 
(version 11, \ncite{xspec}) and assessed the goodness of fit using the 
$\chi^2$ statistic (\ncite{chisquaredstat}). 

As described in Sect.\,\ref{sec:spectra} we merged the SIS0 and SIS1
spectra and also the GIS2 and GIS3 spectra, and then binned both the
resulting spectra such that there were at least twenty
counts in each bin. We fitted the SIS and GIS spectra simultaneously, but
allowed for a constant offset in their normalisation to take into
account calibration uncertainties. 

\subsection{Simple model}
\label{sec:simple}
We began by fitting all our spectra with a single-temperature
optically-thin thermal
plasma model (the {\it mekal} model, 
developed by \ncite{mewe86} and \ncite{mekalfel}) absorbed by
photo-electric absorption by neutral material (the {\it wabs} model, 
using the cross-sections of \ncite{morrison83}). We used the abundances of \scite{anders89}.

\begin{table*}
  \caption{Results of fitting all our \asca\ spectra with our simple
  model (see Sect.\,\ref{sec:simple}). All errors refer to
  $\Delta\chi^2$=2.71 (90 per cent confidence for one parameter of
  interest).  Observations listed in bold face are not fitted with more
  complex models and their observed \asca\ GIS fluxes are presented here. 
  Correcting fluxes for interstellar absorption would have only a small 
  effect in this energy range (typically a few per cent).}
\begin{center}
\begin{tabular}{lcrcccrrccc}
 & Optical & GIS1+2 & $n_{\rm H}$ & Temperature & 0.8--10\,keV Flux & & & & & Dominant\\
Target & State & counts & [$10^{20}$ cm$^{-2}$] & [keV] & [$\rm erg\,s^{-1}\,cm^{-2}$]& $\chi^2$ & $\nu$ & $\chi^2_\nu$ & $P_{null}$ & residuals\\
\hline
V603 Aql&HS& 40827 & $3.2_{-0.6}^{+0.6}$&$7.2_{-0.2}^{+0.2}$& & 1315.8 & 679 & 1.94 & 10$^{-43}$& $<$2\,keV, Fe\\
TT Ari&HS& 9225& $<0.5$&$7.7_{-0.4}^{+0.5}$& & 535.3 & 463 & 1.16 & 1.1$\times$10$^{-2}$ & $<$1.5\,keV\\
{\bf KR Aur}&HS& 1453& $<1.7$&$4.7_{-0.3}^{+0.6}$& 2.3$\times$10$^{-12}$ & 243.4 & 160 & 1.52 & 2.4$\times$10$^{-5}$ & - \\
{\bf CR Boo}&HS?& 1633& $<1.2$&$3.1_{-0.2}^{+0.2}$& 1.3$\times$10$^{-12}$ & 220.8 & 175 & 1.26 & 1.1$\times$10$^{-2}$ & - \\
Z Cam &OB & 3879 & $<0.3$&$4.1_{-0.3}^{+0.3}$& & 723.4 & 385 &1.88&10$^{-23}$  & $<$1.5\,keV\\
Z Cam &T& 12910 & $31_{-3}^{+4}$&$20_{-5}^{+4}$& &775.4 & 603 & 1.29& 2.4$\times$10$^{-6}$ & $<$1\,keV, Fe\\
OY Car&Q & 2838& $35_{-7}^{+7}$&$10_{-2}^{+3}$& & 241.6& 256 &0.94& 0.73 & - \\
HT Cas&Q & 4331& $37_{-5}^{+5}$&$9.1_{-1.1}^{+1.8}$& & 278.2 & 318 & 0.87 & 0.95 & - \\
V436 Cen &Q & 11881 & $6.6_{-1.8}^{+2.1}$&$7.7_{-0.5}^{+0.6}$& & 585.3 & 548 & 1.07& 0.13 & Fe \\
WW Cet&Q & 12438 & $<0.4$&$6.3_{-0.3}^{+0.3}$& & 644.3 & 513 & 1.26 & 6.8$\times$10$^{-5}$ & Fe\\
{\bf GO Com}&? & 623 & $61_{-48}^{+54}$&$3.6_{-1.3}^{+4.1}$& 1.9$\times$10$^{-12}$& 60.6 & 59 & 1.03 & 0.42 & - \\
GP Com & ? & 2299&$<1.2$&$3.9_{-0.3}^{+0.3}$& & 233.7& 171 & 1.37 & 1.0$\times$10$^{-3}$ & - \\
GP Gom & ? & 7495& $<0.5$&$4.3_{-0.2}^{+0.2}$& & 681.5 & 413 &1.65& 1.7$\times$10$^{-15}$ & $<$1.5\,keV, Fe\\
{\bf EY Cyg}&Q &914 &$28_{-25}^{+44}$&$>10$& $3.1\times10^{-12}$& 180.9& 104 & 1.74 & 4.5$\times$10$^{-6}$ & - \\
SS Cyg  &OB& 42346 & $>0.01$&$4.2_{-0.1}^{+0.1}$& & 5112 & 761 & 6.72 & 0 & $<$2\,keV, Fe\\
SS Cyg &Q & 14270&$<0.3$&$8.9_{-0.4}^{+0.4}$& & 561.5 & 573 & 0.98 & 0.63 & Fe \\
U Gem &Q & 14853 &$<0.7$&$8.8_{-0.4}^{+0.4}$& & 649.5 & 583 & 1.11 & 2.9$\times$10$^{-2}$ & - \\
U Gem &Q & 15004 &$1.7_{-1.5}^{+1.6}$&$10.1_{-0.81}^{+0.86}$& & 614.4 & 593 & 1.04 & 0.26 & $<$1.5\,keV\\
VW Hyi &Q & 3764&$<0.6$&$3.2_{-0.2}^{+0.2}$ && 327.4 & 249 & 1.31 & 4.8$\times$10$^{-4}$ & - \\
{\bf VW Hyi} &Q & 354 & $<5.2$&$4.0_{-0.6}^{+0.9}$ & $1.7\times10^{-12}$ & 32.6 & 36 & 0.91 & 0.63 & - \\
T Leo &Q & 11264 & $<1.1$&$4.6_{-0.1}^{+0.2}$ && 541.1& 471 & 1.15 & 1.3$\times$10$^{-2}$ & -\\
V426 Oph&Q & 13497 & $94_{-3}^{+4}$& $>74$ & & 1288.5& 666 & 1.93 & 10$^{-42}$ & $<$2\,keV, Fe \\
LS Peg & ? & 977 & $54_{-13}^{+15}$& $>55$ &  & 282.8& 131& 2.16 &3.5$\times$10$^{-13}$ & $<$2\,keV, Fe\\
RU Peg&OB& 2875& $<0.3$&$4.4_{-0.3}^{+0.3}$ & & 412.7& 202 &2.04& 1.4$\times$10$^{-16}$ & $<$1.5\,keV\\
CP Pup&HS& 3030& $<7$ & $>23$ &  & 188.6 & 193 & 0.98 & 0.58 & - \\
WZ Sge&Q & 6267& $5.4_{-2.1}^{+3.0}$&$4.9_{-0.4}^{+0.4}$ & & 554.8& 395& 1.40 &1.8$\times$10$^{-7}$ & Fe \\
EI UMa& ? & 16054 & $12.5_{-1.5}^{+1.5}$& $>77$ & &752.5 &663 & 1.14 & 8.8$\times$10$^{-3}$ & $<$2\,keV, Fe \\
{\bf SU UMa}&OB& 1927& $<0.8$ & $5.1^{+0.5}_{-0.4}$ & $3.8\times10^{-12}$& 205.9 & 172& 1.20 & 4.0$\times$10$^{-2}$ & $<$1\,keV \\
{\bf CU Vel}&Q & 1894& $<6.0$ & $3.5^{+0.6}_{-0.5}$ & $1.1\times10^{-12}$ & 139.0 & 147 & 0.95 & 0.67& - \\
IX Vel&HS& 5293& $<0.2$ & $4.3^{+0.2}_{-0.2}$ & & 615.9& 336 & 1.83& 9.8$\times$10$^{-19}$ &  $<$1.5\,keV\\\hline
\end{tabular}
\end{center}
\label{table:wabsmekal}
\end{table*}

\begin{figure}
\centering
\includegraphics[height=8.4cm,width=6cm,angle=270]{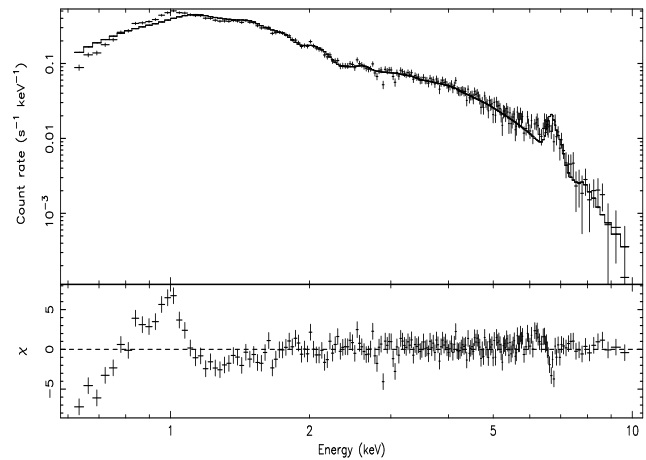}
\caption{\asca\ SIS spectrum of V603~Aql fitted with our simple model. It
can be seen that the model fails to reproduce the spectrum at low
energies and around the iron K-shell complex. 
}
\label{fig:poorfit}
\end{figure}

This simple model resulted in statistically acceptable fits to 14 of 
the 30 spectra 
(a null hypothesis probability of $P_{null}>10^{-2}$). 
An example of typical residuals for the poorly fitted systems is plotted in 
Fig.\,\ref{fig:poorfit}. 
Table\,\ref{table:wabsmekal} shows the fitted parameters for all our
spectra. 
Systems listed in bold face were not
fitted subsequently with more complex models. 


\subsection{Complex model}
\label{sec:complex}

Spectra that were poorly fit with our simple model tended to exhibit residuals at low energies (less than $\sim$1.5\,keV) and around the iron K-shell complex at 6.4-7.0\,keV.  In order to model these spectra correctly, we have attempted to fit the spectra with a more complex model (this has only been attempted on those spectra with at least 2000 counts, with the exception of LS~Peg due to the exceptionally poor fit of $\chi^2_\nu>2$ with the simple model).

Residuals at low energies (e.g.\ Fig.\,\ref{fig:poorfit})
may be the result of inadequate modeling of photo-electric 
absorption in the spectrum. More complex absorption models may be
appropriate if the absorption is dominated by material in the system
itself, in which case the absorbing medium may be clumpy and/or
partially ionised. Both effects tend to allow soft photons to
leak through the absorber, and we found we could not distinguish 
between these effects at the spectral resolution of \asca.
To allow for absorption in our complex model we followed the example of
\scite{do97} and allowed the absorber to be partially ionised. To do
this we employed the {\it absori} model in {\it XSPEC} 
\cite{done92,zdziarskiabsori}. 
We fitted the {\it absori} model with two free parameters: 
the column density of the absorbing medium, and the ionisation parameter $\xi=L/nR^2$, 
where $L$ is the integrated source luminosity between 5\,eV and
300\,keV, $n$ is the density of the material, and $R$ is the distance
of the material from the illuminating source (the temperature of the absorber was fixed at $5\times10^{4}$\,K in all cases).
For systems with interstellar absorption densities known precisely 
from non X-ray methods
(listed in Table\,\ref{table:done}) we also included neutral absorption
fixed at this value. 

Residuals around the iron lines (e.g.\ Fig.\,\ref{fig:poorfit}) tend to
be a sign of fluorescence from cold material (X-ray reflection) 
and/or an inadequate 
representation of the temperature structure of the spectrum. 
With only one free temperature the fitting process
tends to find a best fit to the continuum (which contributes to all data
points) and to allow a poor fit to the emission lines (which
contribute only to a limited number of data points). 
In a cataclysmic
variable we expect emission from a range of temperatures as the
shock-heated gas cools to settle onto the surface of the white dwarf
\egcite{wheatley96a,do97,Mukai03}. 
The temperature distribution
can also effect the fit at low energies,
particularly around the iron L-shell complex ($\sim$1\,keV).

To allow for a range of temperatures we employed the 
{\it cevmkl} model in {\it XSPEC}, which is a multi-temperature 
plasma emission model based on the {\it mekal} model.  
The emission measure follows a power-law in temperature, 
proportional to $(T/T_{\rm max})^\alpha$. 
We chose this model because it allowed us to fit the full range of outburst 
and quiescent spectra with a single simple model. Cooling flow models are 
often a good representation of quiescent spectra \egcite{Mukai03} but not 
outburst spectra. 

We allowed $\alpha$ and $T_{\rm max}$ to be free parameters, 
but found that $T_{\rm max}$ typically favoured high values that are 
 poorly constrained with \asca\ data. To avoid unphysical models
we limited $T_{\rm max}$ to 20\,keV. 
We also included a narrow emission line in our model fixed at 6.4\,keV
in order to account for any fluorescent emission from neutral iron.

\begin{table*}
\begin{center}
\caption{Results of fitting our \asca\ spectra with our more complex
model (see Sect.\,\ref{sec:complex}). Errors refer to
$\Delta\chi^2$=2.71 (90 per cent confidence for one parameter of
interest).   The 0.8--10\,keV flux is the observed GIS flux. 
Correction for interstellar absorption would have little effect on 
the flux in this energy range (just four per cent for SS~Cyg).  
Abundances were fixed at solar values.  The final column gives the F-test probability of this improvement (over the simple model) occuring by chance, showing that there is an significant improvement in all but 1 observation.}
\label{table:done}
\begin{tabular}{l@{\hspace{-0.1cm}}cl@{\hspace{-0.1cm}}cc@{\hspace{-0.1cm}}crrrccclr}
 & Optical & fixed $n_{\rm H}$ &  & free $n_{\rm H}$  &  & 0.8--10\,keV flux & & & && F-test\\
Target & State & {\small [10$^{20}$ cm$^{-2}$]} & $\alpha$ & [10$^{20}$ cm$^{-2}$] & $\xi$ & [$\rm ergs\,s^{-1}\,cm^{-2}$] & $\chi^2$ & $\nu$ & $\chi^2_\nu$ & $P_{null}$ & prob.\\
\hline
V603 Aql& HS&& $0.64^{+0.07}_{-0.06}$ & $30^{+5}_{-5}$ & $0.03^{+0.04}_{-0.02}$ & 30.6$\times$10$^{-12}$ & 744.1 & 677 & 1.10 & 0.04 & $<10^{-12}$\\
TT Ari& HS&& $0.25^{+0.09}_{-0.11}$ & $65^{+12}_{-5}$ & $0.16^{+0.26}_{-0.11}$ & 15.4$\times10^{-12}$ & 433.9 & 461 & 0.94 & 0.81 & $<10^{-12}$ \\
Z Cam  & OB & 0.4$^a$ & $-0.06^{+0.11}_{-0.16}$ & $80^{+36}_{-27}$ & $9.5^{+6.6}_{-4.0}$ & 1.3$\times10^{-12}$ & 416.2 & 383 & 1.09 & 0.12 &  $<10^{-12}$\\
Z Cam  & T & 0.4 & $1.49^{+0.23}_{-0.26}$ & $93^{+13}_{-11}$ & $15.3^{+6.5}_{-5.2}$ & 22.7$\times10^{-12}$ & 636.0 & 601 & 1.06 & 0.16 &  $<10^{-12}$\\
OY Car& Q & & $1.06^{+0.58}_{-0.43}$ & $79^{+29}_{-23}$ & $<20.0$ & 3.0$\times10^{-12}$ & 227.6 & 254 & 0.90 & 0.88 & 10$^{-7}$ \\ 
HT Cas & Q && $1.07^{+0.48}_{-0.70}$ & $47^{+26}_{-11}$ & $<0.04$ & $4.7\times10^{-12}$ & 276.6 & 316 & 0.88 & 0.95 & 0.16 \\
V436 Cen& Q && $0.73^{+0.23}_{-0.16}$ & $28^{+6}_{-1}$ & $<0.001$ & $18.2\times10^{-12}$ & 562.2 & 546 & 1.03 & 0.31 & 10$^{-10}$\\
WW Cet& Q && $0.04^{+0.14}_{-0.10}$ & $69^{+1}_{-2}$ & $<0.02$ & $15.9\times10^{-12}$ & 604.2 & 511 & 1.18 & 0.003 & $<10^{-12}$\\
GP Com& ? && $-0.21^{+0.24}_{-0.14}$ & $65^{+9}_{-8}$ & $<0.02$ & $5.5\times10^{-12}$ & 202.3 & 169 & 1.20 & 0.04 & 10$^{-10}$\\
GP Com & ? && $0.24^{+0.10}_{-0.14}$ & $30^{+2}_{-1}$ & $<0.001$ & $6.7\times10^{-12}$ & 518.6 & 411 & 1.26 & $10^{-4}$ &  $<10^{-12}$\\
SS Cyg & OB& 0.35$^b$ & $0.28^{+0.07}_{-0.06}$ & $31^{+6}_{-7}$ & $1.85^{+0.27}_{-0.22}$ & $40.8\times10^{-12}$ & 986.6 & 759 & 1.30 & $10^{-8}$ &  $<10^{-12}$ \\ 
SS Cyg & Q & 0.35     & $1.48^{+0.07}_{-0.06}$ & - & - & $22.6\times10^{-12}$ & 520.7 & 571 & 0.91 & 0.94 & $<10^{-12}$ \\
U Gem & Q & 0.31$^c$ & $0.99^{+0.16}_{-0.27}$ & $16^{+14}_{-6}$ & $<1$ & $15.9\times10^{-12}$ & 614.1 & 581 & 1.06 & 0.17 &  $<10^{-12}$ \\
U Gem & Q & 0.31     & $1.03^{+0.24}_{-0.18}$ & $18^{+8}_{-9}$ & $<0.08$ & $17.7\times10^{-12}$ & 568.6 & 591 & 0.96 & 0.74 & $<10^{-12}$ \\
VW Hyi & Q & 0.006$^d$ &$1.74_{-0.61}^{+0.43}$ & $<16$ & - & $4.7\times10^{-12}$ & 272.5 & 246 & 1.11 & 0.12 & $<10^{-12}$ \\
T Leo & Q &&$0.002_{-0.071}^{+0.089}$ & $47^{+3}_{-7}$ & $<0.02$ & $11.1\times10^{-12}$ & 496.0&469 & 1.06 & 0.19 &  $<10^{-12}$ \\
V426 Oph& Q &&$2.51_{-0.58}^{+0.63}$ & $350_{-25}^{+23}$ &$73_{-9}^{+8}$ & $35.9\times10^{-12}$ & 742.6 & 664 & 1.12 & 0.02 & $<10^{-12}$\\
LS Peg& ? && $1.21^{+0.41}_{-0.53}$ & $650_{-120}^{+180}$ &$210_{-60}^{+74}$ & $1.9\times10^{-12}$ &188.1 &130 &1.45 & 10$^{-3}$ & $<10^{-12}$ \\
RU Peg& OB&&$0.26_{-0.19}^{+0.10}$ & $32_{-17}^{+28}$ & $0.45_{-0.44}^{+0.62}$ & $4.8\times10^{-12}$ & 222.0& 200 & 1.11 & 0.14&  $<10^{-12}$\\
CP Pup & HS& &$1.10_{-0.48}^{+0.98}$& $150_{-127}^{+124}$ & $112_{-87}^{+272}$ & $4.7\times10^{-12}$ & 182.4 & 191 & 0.95 & 0.66 & 10$^{-3}$ \\ 
WZ Sge& Q &&$1.29_{-0.31}^{+0.55}$ & $44_{-21}^{+16}$ & $2.0^{+1.5}_{-1.1}$ & $2.5\times10^{-12}$ & 531.9 & 392 & 1.36 & $10^{-6}$ &  10$^{-10}$ \\
EI UMa & ? && $>9.2$ & $124^{+23}_{-18}$ & $111_{-25}^{+31}$ & $30.0\times10^{-12}$ & 627.6 & 661 & 0.95 & 0.82 &  $<10^{-12}$\\
IX Vel& HS&0.20$^a$&$-0.01_{-0.22}^{+0.11}$ & $78_{-17}^{+18}$ & $1.7_{-1.5}^{+4.2}$ & $8.3\times10^{-12}$ & 374.3& 333&1.12 & 0.06 &  $<10^{-12}$\\

\hline
\end{tabular}
\end{center}
$^a$ C. Mauche, 2000, private communication \\
$^b$ \scite{nh88} \\
$^c$ \scite{ugemeuve} \\
$^d$ \scite{nh90} \\
\end{table*}

\begin{figure}
\centering
\includegraphics[height=8.4cm,width=6cm,angle=270]{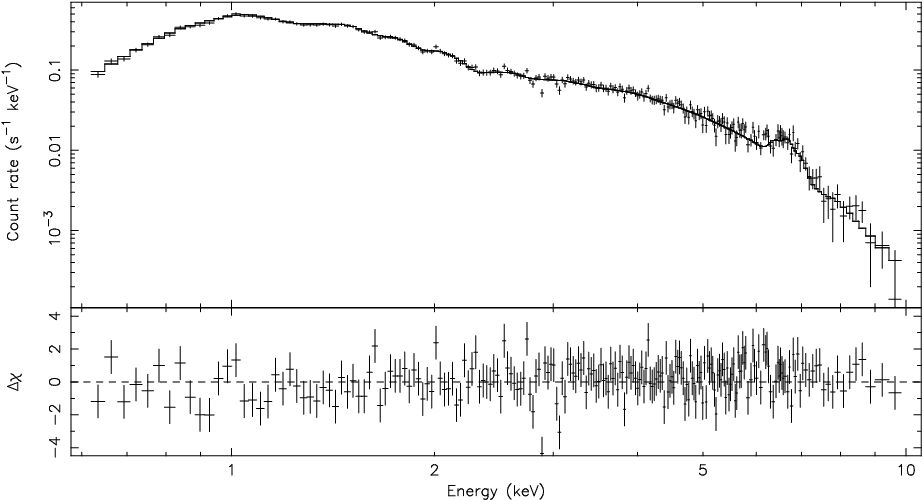}
\includegraphics[height=8.4cm,width=6cm,angle=270]{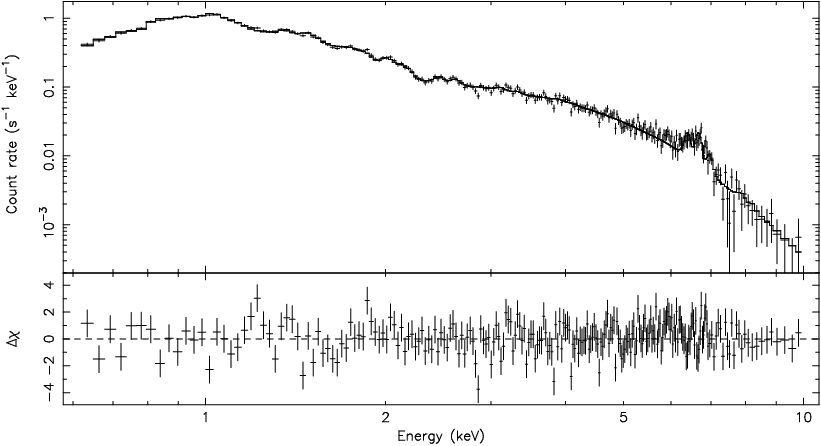}
\caption{\asca\ SIS spectrum of our brightest two systems, V603~Aql (top)
and SS~Cyg (bottom), fitted with our more complex model and with free
metal abundance (see text). 
}
\label{fig:goodfit}
\end{figure}

Fitting with our more complex model yielded acceptable fits to all the observations with only 4 exceptions; GP~Com, SS~Cyg (during outburst), LS~Peg, and WZ~Sge.  
Three of these were well fit once we allowed abundances to vary as a free parameter (Sect.\,\ref{sec:abund}). 
Although almost half of the spectra were well fitted with the simple model, all but one of the 30 spectra are significantly better fit with our complex model.
Fitted parameters for all our fits with our complex model are
presented in Table\,\ref{table:done}.
Parameters for the free abundance fits are presented in Table\,\ref{table:donefreeabund}.
In Fig.\,\ref{fig:goodfit} we show our best fits to the \asca\ spectra
of our two brightest systems SS~Cyg and V603~Aql (including the
modified abundances described below). 

\subsubsection{Temperature distribution}
\begin{figure}
\centering
\includegraphics[width=8.4cm]{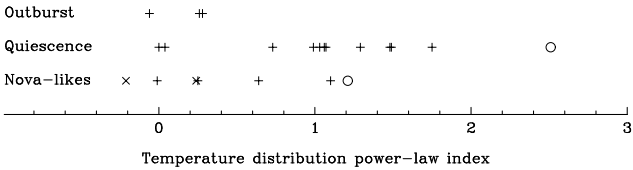}
\caption{Fitted temperature-distribution power-law indices ($\alpha$)
for our complex model (see text). The crosses indicate AM CVn systems.  
The two systems with extreme GIS hardness ratios in our sample,
V426 Oph and  LS~Peg, are indicated with circles. 
}
\label{fig:alphas}
\end{figure}

The fitted $\alpha$ values in Table\,\ref{table:done} exhibit a broad
range, indicating very different temperature distributions in 
different systems. 
SS~Cyg and Z~Cam were observed both in outburst and quiescence, 
and it is clear that the temperature distribution changed
dramatically between the two states. In outburst there is a much
larger contribution from cool gas. Indeed, this difference seems to
account for the universally low hardness ratios of dwarf novae in
outburst (see Fig.\,\ref{fig:ratiostates} and
Sect.\,\ref{sec:ratio}). 
In Fig.\,\ref{fig:alphas} we plot the $\alpha$ values by system type
and state. It can be seen that the dwarf novae in outburst have a much 
larger contribution from cool gas than most in quiescence. 
There is also a surprisingly wide range of temperature distributions in 
the quiescent spectra. The harder spectra are consistent with cooling flow 
spectra, but many are much softer. 
The nova-like variables 
seem to
have temperature distributions that are
intermediate between dwarf novae in outburst and
quiescence. Our results certainly support the conclusions of
\scite{Mauche02} who show that the X-ray spectra of nova-like
variables are not black-bodies. 

The maximum temperature, $T_{\rm max}$, is poorly constrained in our fits, 
and values below 20\,keV
were favoured in only four cases: the earlier observation 
of VW~Hyi (5\,keV), WZ~Sge (7\,keV), the later  
GP~Com observation (14\,keV), and IX~Vel (12\,keV).

\subsubsection{Absorption}
The absorption column density in our complex model fits 
(Table\,\ref{table:done}) is uniformly higher than in the
single-temperature fits (Table\,\ref{table:wabsmekal}). 
This can be understood as a natural result of
the range of temperatures in the X-ray spectra of non-magnetic
cataclysmic variables. In the single-temperature fits the model is
underestimating the quantity of cool gas in all cases, and so
the fitting process attempts to compensate by minimising absorption
(which effects only the soft band). Once a realistic range of
temperatures is included we see that the fit is free to increase the 
amount of absorption. 

In some cases, i.e. V426~Oph, LS~Peg and
 EI~UMa, it is clear that strong absorption is required. 
These three systems also require the absorption to be ionised (or
clumpy, or time variable). 
In all other systems there is no requirement for the absorbers to be
strongly ionised. 

\begin{figure}
\centering
\includegraphics[height=8.4cm,angle=270]{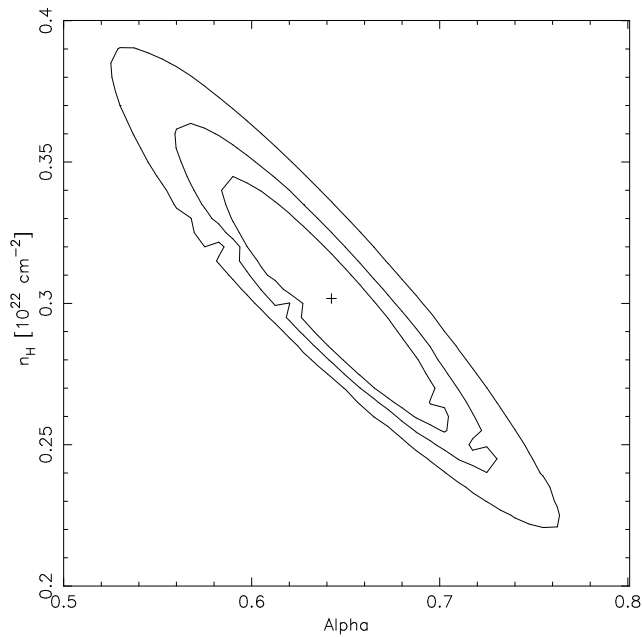}
\caption{Allowed ranges of $n_{\rm H}$  and $\alpha$ from  fitting our
model of Sect.\,\ref{sec:complex} to the \asca\ spectra of V603~Aql.  
The cross represents the best fit values. The three contours represent
$\Delta\chi$ of 2.3, 4.61 and 9.21, corresponding to 68, 90 and 99 per
cent confidence for two parameters of interest \protect\cite{chisquaredstat}.
It can be seen that these fitted quantities 
are strongly correlated. 
}
\label{fig:confidence}
\end{figure}

We find that there is a tendency for the
absorption column density to play off against the temperature
distribution.  As an example, we plot confidence
contours for the fitted values of $n_{\rm H}$ and $\alpha$ from the
spectrum of V603~Aql in Fig.\,\ref{fig:confidence}. 
It can be seen that the parameters are correlated. 
Therefore, if the contribution from cool gas is overestimated in our
models, this might lead to a systematic overestimate of the absorption
column density. We believe this is quite likely, since our power law
temperature distribution does indeed have a larger contribution from
cool gas than do cooling flow models (in which the emission measure is
weighted by the inverse of the emissivity at that temperature). 
For this reason we advise caution in interpreting our fitted column
densities. 

In order to asses whether intrinsic absorption is required at all (in
systems other than V426~Oph, LS~Peg 
and EI~UMa) we
turn to the subset of systems for which we have an independent
estimate of the interstellar absorption column densities (indicated in
Table\,\ref{table:done}). 
Using the F-test statistic we find that inclusion of absorption in
excess of the interstellar value does not make a significant
contribution to our fits for the quiescent dwarf novae 
(SS~Cyg and VW~Hyi). However, the improvement in the fit statistic {\em
is} significant (at $>$99.9\% confidence) 
in both dwarf novae in outburst (Z~Cam and SS~Cyg) and
also in the nova-like IX~Vel. 
The only exception to this rule is U~Gem, where there is clear
evidence for orbital-phase dependent intrinsic absorption dips in the
\asca\ lightcurve (see Sect.\,\ref{sec:orbitalmodulations} and 
Fig.\,\ref{fig:orbital}).

\subsubsection{X-ray reflection}
\label{sec:reflection}
The 6.4\,keV line contributed significantly to our fits  
in four cases: V436~Cen, SS~Cyg (in outburst), V426~Oph and EI~UMa. The best fitting 
equivalent widths are 170$\pm$50, 95$\pm$25, 185$\pm$40 and 200$\pm$45\,eV 
respectively. 
Despite these detected fluorescent lines, an X-ray reflection
continuum was not required in order to achieve an statistically acceptable fit 
in any of our spectra. 
This is probably because the effective area 
of \asca\ dropped steeply above 10\,keV,  
where the reflection continuum becomes increasingly important. 
We note, however, that inclusion of a reflection continuum would
act to reduce the required temperatures \egcite{do97}, and possibly
the absorption column density also. 
As an example we applied the reflection continuum model from
\scite{do97} to the very hard spectrum of EI~UMa. Without a reflection continuum 
EI~UMa favoured very high values of $\alpha>$9.2. However, with the
reflection continuum added we found an equally good fit with a value 
$\alpha=$0.45$^{+0.06}_{-0.08}$. We also found that the equivalent width of
the 6.4\,keV line dropped from 200\,eV to $<$56\,eV. None of our other
spectra could constrain the reflection continuum model, but we note that
such a component probably makes a significant contribution in other
systems, and would tend also to reduce the fitted temperatures. 

\subsection{Abundances}
\label{sec:abund}
  
  The spectral fitting using our complex model still resulted in unacceptable reduced chi-squared values for four observations (GP~Com, SS~Cyg in outburst, LS~Peg and WZ~Sge). In an attempt to improve these fits we allowed the abundances to vary as a free parameter. This resulted in statistically acceptable fits to three of these four remaining spectra. The poor fit in the fourth system, LS~Peg, seems to be due to a random statistical fluctuation in a few GIS spectral bins. The SIS spectrum is well fit with our model.

\begin{figure}
\centering
\scalebox{0.5}{\includegraphics{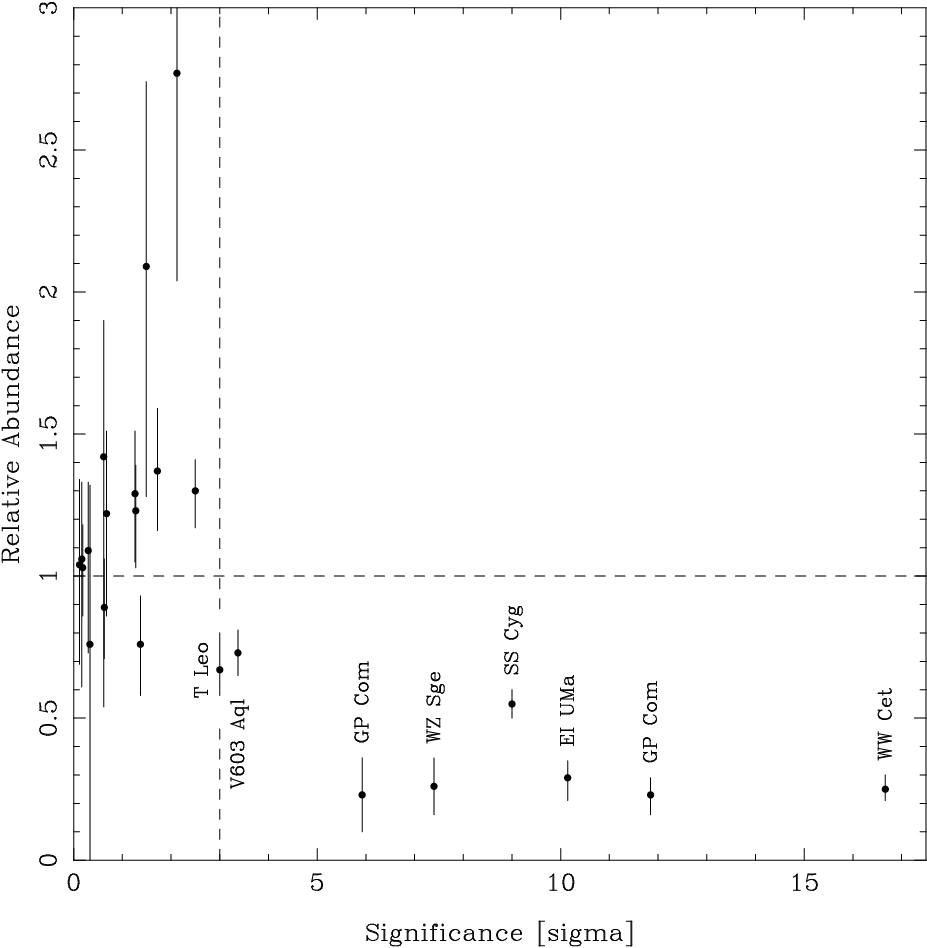}}
\caption{Plot of the metal abundance (relative to solar) against significance [$|A-1.0|/error$], taken from the fitting of our more complex model of Sect.\,\ref{sec:complex}.  Those targets with a significance greater than or equal to 3 (i.e. those with errors less than a third the size of the deviation from solar abundance) have been labelled.}
\label{fig:donefreeabund}
\end{figure}

We have also allowed the abundances to vary for those spectra that give statistically acceptable fits with solar abundances, and have accessessed improvements in the fits using the F-test (see Table~\ref{table:donefreeabund}).  We found significant improvements via the F-test in 10 of the observations. 
  Figure~\ref{fig:donefreeabund} shows the fitted metal abundances (relative to the solar abundances of \ncite{anders89}) against the significance of the deviation from solar abundance (calculated using $|A-1.0|/error$) for all the observations requiring the complex modelling of Sect.~\ref{sec:complex}.  All significant deviations from solar abundance are to sub-solar values. 


\begin{table*}
\begin{center}
\caption{Results of fitting our \asca\ spectra with our more complex
model (see Sect.\,\ref{sec:complex}) and free metal abundances. Errors refer to
$\Delta\chi^2$=2.71 (90 per cent confidence for one parameter of
interest). A reflection continuum has also been included in the model for 
EI~UMa. }
\label{table:donefreeabund}
\begin{tabular}{lccccccccc}
  &  & & free $n_{\rm H}$  & & & & & \multicolumn{2}{c}{F-Test}\\
Target & Abundance & $\alpha$ & [10$^{20}$ cm$^{-2}$] & $\xi$ & $\chi^2$ & $\nu$ & $\chi^2_\nu$ & {\small Statistic} & {\small Probability}\\
\hline
V603 Aql & $0.73^{+0.08}_{-0.08}$ & $0.55^{+0.07}_{-0.06}$ & $29^{+5}_{-4}$ & $0.02^{+0.04}_{-0.01}$ & 715.6 & 676&1.06 & 27 & 10$^{-7}$ \\ 
TT Ari   & $0.89^{+0.18}_{-0.17}$ & $0.22^{+0.10}_{-0.10}$ & $64^{+7}_{-12}$ & $0.17^{+0.35}_{-0.11}$ & 432.6& 460& 0.94& 1.4 & 0.2 \\ 
Z Cam & $1.29^{+0.24}_{-0.22}$ & $1.46^{+0.29}_{-0.23}$ & $90^{+12}_{-11}$ & $12.6^{+6.6}_{-4.8}$ & 631.5& 600& 1.05& 4.3 & 0.04\\
Z Cam & $1.42^{+0.88}_{-0.48}$ & $0.002^{+0.15}_{-0.08}$ & $83^{+36}_{-26}$ & $8.2^{+5.3}_{-3.6}$ & 414.3& 382& 1.08& 1.6 & 0.2\\
OY Car & $2.09^{+0.81}_{-0.65}$ & $1.12^{+0.73}_{-0.43}$ & $70^{+31}_{-14}$ & $<12.4$ & 219.5& 253& 0.87& 9.3 & 0.002\\
HT Cas & $1.04^{+0.35}_{-0.30}$ & $1.08^{+0.48}_{-0.32}$ & $47^{+26}_{-11}$ & $<0.04$ & 276.6& 315& 0.88& 0.07 & 0.8 \\
V436 Cen & $1.30^{+0.13}_{-0.11}$ & $0.82^{+0.24}_{-0.18}$ & $28^{+8}_{-8}$ & $<0.001$ & 556.6& 545& 1.02& 5.5 & 0.02 \\
WW Cet & $0.25^{+0.04}_{-0.05}$ & $0.01^{+0.03}_{-0.02}$ & $36^{+1}_{-1}$ & $<0.001$ & 513.9& 510& 1.01& 90 & 10$^{-19}$\\
GP Com (1)& $0.23^{+0.13}_{-0.13}$ & $-0.23^{+0.21}_{-0.21}$ & $28^{+56}_{-16}$ & $<0.01$ & 178.9& 168& 1.07& 22 & 10$^{-5}$\\
GP Com (2) & $0.23^{+0.07}_{-0.06}$ & $-0.09^{+0.11}_{-0.10}$ & $24^{+5}_{-5}$ & $<0.001$ & 407.6& 410 & 0.99& 110 & 10$^{-23}$\\
SS Cyg 1 & $0.55^{+0.05}_{-0.05}$ & $0.16^{+0.06}_{-0.06}$ & $25^{+5}_{-5}$ & $1.46^{+0.26}_{-0.24}$ & 852.4& 758& 1.12& 120 & 10$^{-25}$\\
SS Cyg 2 & $0.76^{+0.18}_{-0.17}$ & $1.41^{+0.09}_{-0.09}$ & $61^{+235}_{-61}$ & - & 516.2& 570& 0.91& 5.0 & 0.03 \\
U Gem 1 & $1.03^{+0.17}_{-0.15}$ & $0.92^{+0.23}_{-0.19}$ & $20^{+9}_{-7}$ & $<0.49$ & 613.6 & 580& 1.06& 0.5 & 0.5 \\
U Gem 2 & $1.23^{+0.20}_{-0.16}$ & $1.16^{+0.21}_{-0.18}$ & $13^{+12}_{-5}$ & $<0.1$ & 564.8& 590& 0.96& 4.0&0.05\\
VW Hyi (1) & $1.22^{+0.36}_{-0.29}$ & $1.69^{+0.57}_{-0.58}$ & $6^{+14}_{-6}$ & $<80$ &271.0 & 245& 1.11& 1.4 & 0.3\\
T Leo & $0.67^{+0.09}_{-0.13}$ & $-0.04^{+0.08}_{-0.08}$ & $42^{+10}_{-9}$ & $<0.04$ & 488.7& 468& 1.04& 7.0& 0.009\\
V426 Oph & $1.37^{+0.21}_{-0.22}$ & $2.68^{+0.69}_{-0.63}$ & $336^{+27}_{-21}$ & $68.9^{+9.2}_{-8.2}$ & 734.1& 663& 1.11& 7.7 & 0.006\\
LS Peg & $2.77^{+0.73}_{-0.94}$ & $1.26^{+0.44}_{-0.49}$ & $555^{+163}_{-134}$ & $161^{+81}_{-79}$ & 178.8 & 129 & 1.39 & 6.7 & 0.01 \\
RU Peg & $1.06^{+0.45}_{-0.27}$ & $0.25^{+0.22}_{-0.21}$ & $34^{+26}_{-11}$ & $<1.4$ & 221.8& 199& 1.11& 0.2 & 0.7\\
CP Pup & $0.76^{+0.88}_{-0.56}$ & $1.01^{+1.07}_{-0.44}$ & $165^{+165}_{-140}$ & $125^{+295}_{-97}$ & 182.2 & 190 & 0.96 &0.23 & 0.63\\ 
WZ Sge & $0.26^{+0.10}_{-0.10}$ & $1.15^{+0.70}_{-0.68}$ & $15^{+17}_{-7}$ & $<1.2$ & 459.1& 391& 1.17& 62 & 10$^{-14}$\\
EI UMa & $0.29^{+0.08}_{-0.06}$ & $0.11^{+0.08}_{-0.07}$ & $156^{+24}_{-25}$ & $19.4^{+7.2}_{-7.1}$ &666.1 &659 &1.01 & 44 & 10$^{-10}$\\
IX Vel & $1.09^{+0.36}_{-0.24}$ & $0.00^{+0.21}_{-0.24}$ & $80^{+20}_{-17}$ & $1.6^{+3.9}_{-1.4}$ & 373.8& 332& 1.13& 0.4 & 0.5 \\
\hline
\end{tabular}
\end{center}
\end{table*}

\section{X-ray luminosities}
We used our best fitting models 
to estimate total X-ray
luminosities by integrating over a broad energy range
(0.1--100\,keV).
Correcting for the interstellar absorption in
those cases where it is known (Table\,\ref{table:done}) would 
have a negligible effect on our luminosities (4 per cent in the case
of SS~Cyg). 
Given the uncertainty in the amount of absorption in
many of our spectra,  we did not attempt to correct our fluxes 
for the fitted absorption component. 
However, in those systems with well constrained strong absorption components,
V426~Oph 
and EI~UMa,
we note that setting the column density to zero increases the fluxes
presented in Table\,\ref{table:done} by factors of 1.36 
 and 1.09 respectively. 
The presence of reflection continua, which were not constrained in our spectral fitting, may also modify these luminosities slighty.

In most cases the distances to cataclysmic variables are highly
uncertain, but there is now a growing number of systems for which
parallaxes have been measured. Ten systems within our sample have
measured parallaxes (see Table\,\ref{table:lum}) and we place
particular emphasis on these systems. 
Distances for other systems in our sample were taken from many
sources, listed in table\, \ref{table:general}.

\begin{table}
\caption{\asca\ X-ray luminosities of systems with measured
parallaxes. Fluxes have been calculated by integrating our best
fitting models in the range 0.01--100\,keV.
}
\label{table:lum}
\begin{center}
\begin{tabular}{lcc@{\hspace{-0.05cm}}cc}
 & & Distance & Flux & Luminosity\\
Name & State & [pc] & [$\rm erg\,s^{-1}\,cm^{-2}$] & [$\rm erg\,s^{-1}$]\\ \hline
V603 Aql & HS & 237$\pm^{380,a}_{\phantom{3}90}$ & $45\times10^{-12}$ & $3.0\times10^{32}$ \\
Z Cam    & OB & 163$^{+68,b}_{-38}$ & $5\times10^{-12}$ & $1.4\times10^{31}$\\
 & & & $40\times10^{-12}$ & $1.3\times10^{32}$\\
GP Com	 & ? & 68$^{+7,b}_{-6}$ & $7\times10^{-12}$ & $3.8\times10^{30}$\\
 & & & $7\times10^{-12}$ & $3.6\times10^{30}$\\
SS Cyg   & OB & 166.2$\pm12.7^c$ & $82\times10^{-12}$ & $2.5\times10^{32}$\\
   & Q & & $39\times10^{-12}$ & $1.2\times10^{32}$\\
U Gem    & Q & 96$\pm^{5,c}_{4}$ & $25\times10^{-12}$ & $2.8\times10^{31}$\\
& & & $27\times10^{-12}$ & $3.0\times10^{31}$\\
T Leo    & Q & 101$^{+13,b}_{-11}$ & $15\times10^{-12}$ & $1.8\times10^{31}$\\
RU Peg   & OB & 287$\pm^{23,d}_{20}$ & $10\times10^{-12}$ & $9.4\times10^{31}$\\
WZ Sge   & Q & 43.5$\pm0.3^d$ & $4\times10^{-12}$ & $9.9\times10^{29}$\\
SU UMa   & OB & 260$^{+190,b}_{-90}$ & $8\times10^{-12}$ & $6.4\times10^{31}$\\
IX Vel   & HS & 96$\pm^{10,a}_{\phantom{0}8}$ & $27\times10^{-12}$ & $3.0\times10^{31}$ \\
\hline
\end{tabular}
\end{center}
$^a$ \scite{hippdist}\\
$^b$ \scite{thorstensendistances}\\
$^c$ \scite{Harrison00}\\
$^d$ \scite{Harrison03}\\
\end{table}

\begin{figure}
\centering
\scalebox{1.1}{\includegraphics{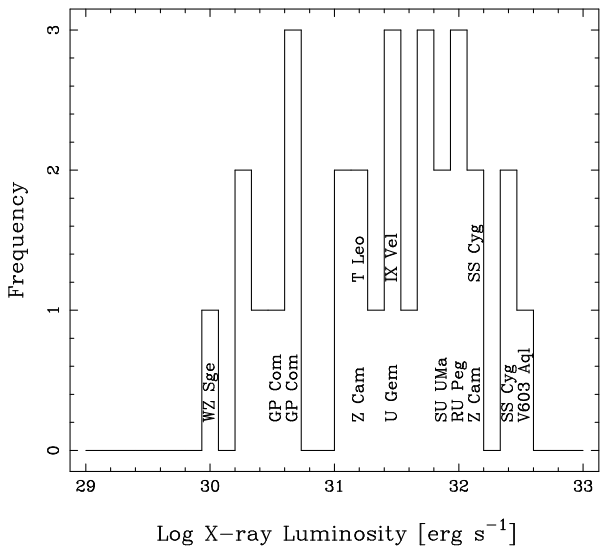}}
\caption{X-ray luminosities of non-magnetic cataclysmic
variables.  
Named systems are those for which parallax measurements are available.
}
\label{fig:lum_hist}
\end{figure}

\begin{figure}
\centering
\scalebox{0.6}{\includegraphics{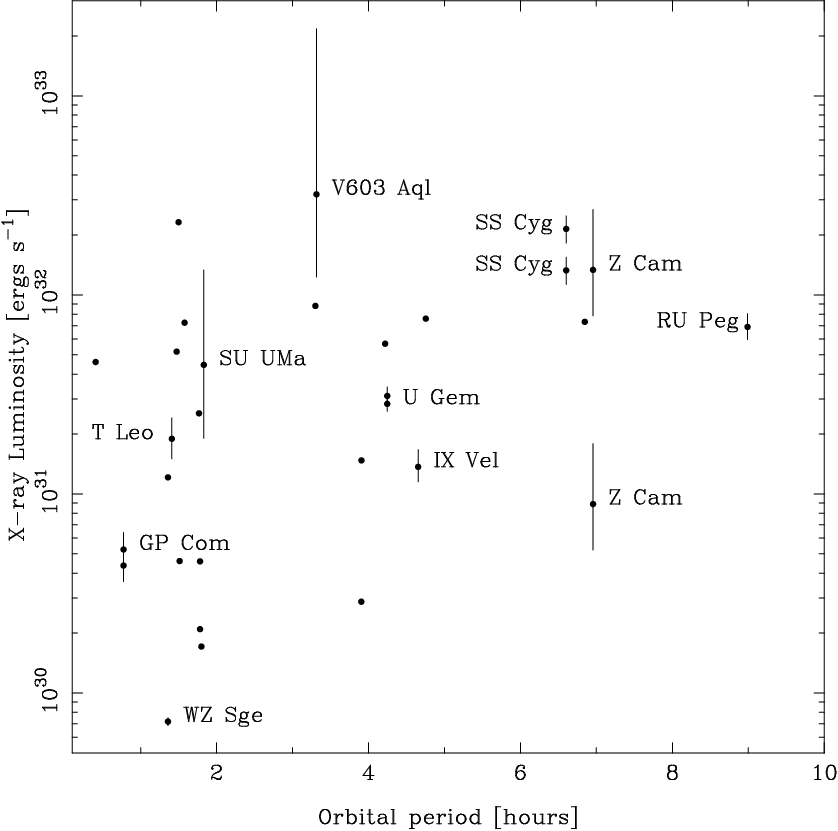}}
\caption{X-ray luminosities of non-magnetic cataclysmic
variables as a function of orbital period. 
Named systems are those for which parallax measurements are available.
}
\label{fig:lum}
\end{figure}

\begin{figure}
\centering
\scalebox{0.6}{\includegraphics{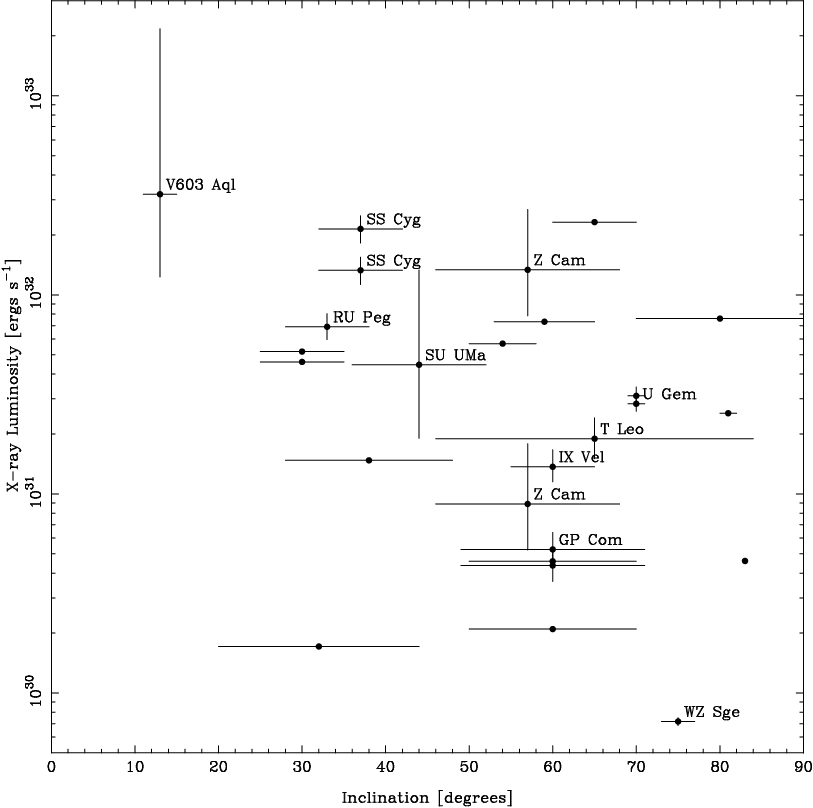}}
\caption{X-ray luminosities of non-magnetic cataclysmic
variables as a function of inclination. 
Named systems are those for which parallax measurements are available.
}
\label{fig:inc}
\end{figure}

X-ray luminosities for the systems with measured parallaxes are
presented in Table\,\ref{table:lum}, and our full set of estimated
luminosities are presented in Fig.\,\ref{fig:lum_hist}. 
They are also plotted as a
function of orbital period in Fig.\,\ref{fig:lum} and inclination in 
Fig.\,\ref{fig:inc}.

Figure\,\ref{fig:lum_hist} shows that hard X-ray luminosities span the range 
$1\times10^{30}\rm\,erg\,s^{-1}$ to $3\times10^{32}\rm\,erg\,s^{-1}$. 
Figure\,\ref{fig:lum} shows there is evidence for a weak
correlation of X-ray luminosity with orbital period, and hence
presumably with long-term mean accretion rate. 
There is also, however, a wide range of luminosities at all orbital periods. 

WZ~Sge is the least luminous  cataclysmic variable in our sample, 
indicating an unusually low accretion rate in quiescence. 
From our X-ray luminosity, we estimate an accretion rate of $6\times10^{12}\rm\,g\,s^{-1}$ using
a white dwarf mass of 1.2\,M$_{\sun}$ \cite{wzsgeprimarymass} and 
radius $5\times10^{8}$\,cm (using the relation of \ncite{whitedwarfradius}).
This low quiescent accretion rate is likely to be related to the extremely long 
outburst recurrence times observed in WZ~Sge. 

In Fig.\,\ref{fig:inc} we plot our X-ray luminosities as a function of
inclination angle.
Our results appear to be consistent with the discovery
of \scite{rosatnmcvs} that X-ray emission in non-magnetic cataclysmic
variables is anti-correlated with inclination. We note,
however, that this figure relies on uncertain inclinations and that the correlation remains
somewhat fragile.

We have also compared the luminosities of this \asca\ sample with the \einstein\ and \rosat\ samples, as previously published by \scite{spectraeinstein} and \scite{rosatcvs} respectively, by extrapolating our best fit spectral models into the spectral ranges of these two satellites and using their estimates for distances.  We find no systematic differences between the samples.
  Large luminosity differences were seen in individual objects (e.g. Z~Cam, V426~Oph and VW~Hyi between the \asca\ and \rosat\ samples, and RU~Peg and SU~UMa between the \asca\ and \einstein\ samples).  These were probably caused by the accretion disk being in a different (optical) state during the two observations.
Smaller differences are also present and may represent variations occurring in the same optical state (as demonstrated by the two \asca\ observations of VW~Hyi). 
  However, some of the differences between the samples will inevitably be caused by the non-overlapping band-passes of the two satellites with \asca\ (especially in the case of \einstein).  The predicted flux below the bandpass of \asca\ is a function of the spectral parameters, in particular the estimated amount of absorption, and may therefore vary significantly from the true flux. 

\section{Discussion and conclusions}

\subsection{X-ray detections and luminosities}
Twenty-five of the 29 systems in our sample were detected with \asca, including four systems that were detected serendipitously.  All of these objects are previously known X-ray sources.
 We find X-ray luminosities in the range 
$1\times10^{30}\rm\,erg\,s^{-1}$ to $3\times10^{32}\rm\,erg\,s^{-1}$. 
In one case, V345~Pav, the X-ray source had previously been misidentified.

\subsection{X-ray variability}
Our timing analysis 
shows that flux variations on timescales
of hours is most commonly associated with the outburst state of dwarf
novae, as expected. We also find, however, that the X-ray flux in
quiescence can differ by up to an order magnitude in individual
systems (e.g. SS~Cyg and VW~Hyi).  
Since X-rays are known to arise close to the
white dwarf, this difference in X-ray flux implies differences in
the accretion rate onto the white dwarf, also by an order of magnitude. 
This points to behaviour in the inner accretion disc that can strongly
effect the accretion rate without triggering an overall state change
in the accretion disc. 
 
Power spectral analysis showed that systems in our sample do not 
typically exhibit periodic modulation of their
X-ray flux. The exceptions were systems viewed at high inclination angles
which tend to exhibit modulation at the orbital period. 
In some cases this modulation is anti-correlated with the hardness
ratio (OY~Car and U~Gem) but in others this correlation is either
absent or reversed (HT~Cas, V436~Cen and T~Leo). We show that the
orbital dips observed in the observations of U~Gem are broadly
consistent with photo-electric absorption. 

Candidate periodicities at other than the orbital period were found
in four systems: WW~Cet 9.85$\pm0.05$\,min, V426~Oph
29.2$\pm$0.9\,min, LS~Peg 29.6$\pm$1.8\,min, and EI~UMa
12.36$\pm$0.09\,min.  
Three of these (V426~Oph, LS~Peg, and EI~UMa) also exhibit distinct X-ray
spectra that lead us to believe these systems are magnetic accretors
(see Sect.\,\ref{sec:ips}). 

\subsection{X-ray spectra}
We found acceptable fits to all the \asca\  spectra 
using thermal plasma models. However, the high 
spectral resolution of the \asca\ CCD detectors revealed complexity in
the X-ray spectra of our brighter systems. Bright systems tended to
require a range of X-ray temperatures, and we found that dwarf novae
in outburst required this distribution to be weighted more to low
temperatures than dwarf novae in quiescence. 
Nova-like variables may have spectra that are intermediate between
quiescent  and outbursting dwarf novae.  Quiescent dwarf novae have a suprisingly wide range of temperature distributions, and many are too soft to be fitted 
with cooling flow models. 

We also found that our spectra of dwarf novae in outburst required
photo-electric absorption in excess of the interstellar value, whereas
absorption in dwarf novae in quiescence was consistent with the
interstellar value.  There was no requirement for these absorbers to be
strongly ionised, except in the two systems discussed in more detail
below (V426~Oph and EI~UMa). 

Outburst absorption is most likely due to accretion disc winds also
seen in the ultraviolet \egcite{Mauche87,knigge97}. 
These disc winds are believed to be only present in outburst and our 
observations support this view. 
However, we note that the absorption is present very early in the
outburst of Z~Cam, during the transition observation (\ncite{baskill01}).

Our brighter systems also tended to favour the inclusion of an
emission line 6.4\,keV, the energy of K-shell emission from neutral
iron. This line is believed to arise from fluorescence of 
relatively cold material that is illuminated by the X-ray
source. Its detection implies the presence also of a reflection
continuum  component that will become increasingly important
at higher energies. Unfortunately the \asca\  spectra do not extend to
sufficiently high energies to constrain the reflection continuum, and
so we did not fit our spectra with such a model. We note that
inclusion of a reflection continuum would act to reduce the
contribution of high temperatures in our fits.

  Significant improvements in the spectral fit were seen for seven spectra when the metal abundance was allowed to vary. All seven favour sub-solar abundances relative to the values of \scite{anders89}. 

It is clear that the increased spectral resolution of \asca\ over
previous instruments has allowed us to detect complexity in the
spectra of non-magnetic cataclysmic variables. 
In instruments such as \heao , \einstein\  and \exosat\  the spectra
were well fit with simple single-temperature emission 
components \egcite{heao1,einsteincvsample,spectraeinstein,exosatME}. 
With \rosat\  it was possible to show that the emission was not from
an isothermal gas, but not to constrain the range of temperatures
\egcite{Vrtilek94,10cvs,richman96}, although it was possible to
constrain multi-temperature models by combining \rosat\  and \ginga\
data \cite{wheatley96a}.
With the CCD resolution of \asca\  we have been able to constrain 
multi-temperature emission models and ionised absorption models, and have
measured sub-solar abundances and found evidence for X-ray
reflection. 

Further progress in this field, however,  relies on the detection of individual
spectral features (emission lines and absorption edges) in order to
break the degeneracy between the temperature distribution and any
ionised, clumpy or variable absorption component. 
Observations with dispersive X-ray
spectrographs are already beginning to do that \cite{ugemchandra,Mukai03,Mauche03,Perna03} but such
observations must have high statistical quality in order to resolve
features such as ionised absorption edges. Effective area at high
energies ($>$10\,keV) is also required in order to constrain the
reflection continuum. 

\subsection{An intermediate polar classification for V426~Oph, LS~Peg and EI~UMa}
\label{sec:ips}
In the hardness ratio plots of 
Figs.\,\ref{fig:hardnessfreqdist}\,\&\,\ref{fig:ratiostates} two systems
stand out with exceptionally hard spectra: V426~Oph and
LS~Peg. Inspection of their best fitting spectral parameters in Tables\,\ref{table:wabsmekal}, \ref{table:done} and \ref{table:donefreeabund}
shows that both are unusual in requiring strong absorption.
V426~Oph also requires the X-ray emission to be
weighted to high temperatures (i.e. a high value of
$\alpha$). 
Further inspection of Tables\,\ref{table:done} and \ref{table:donefreeabund} 
reveals that EI~UMa also requires strong absorption.

These spectral properties are very similar to those observed in
intermediate polar systems \egcite{ascaMCVs}. Strong X-ray absorption
is a common feature of intermediate polars because the magnetic field
of the white dwarf diverts the accretion flow out of the orbital
plane, resulting in large column densities of cool pre-shock gas in
our line of sight to the hot post-shock X-ray emitting region
\egcite{Rosen88,Buckley89}. 

Remarkably, 
V426~Oph, LS~Peg and EI~UMa
also account for three of the four
candidate periodicities identified in 
Sect.\,\ref{sec:shortperiodmodulations}. Detection of a
coherent X-ray period other than the orbital period is the generally
accepted criteria for identification of a system as an intermediate
polar \egcite{patterson94}. 

We argue that the distinct X-ray spectra of V426~Oph,
LS~Peg and EI~UMa represent excellent evidence for
magnetically-confined accretion
in these systems. The detection of candidate spin periods provides 
strong supporting evidence, and we suggest that these systems be
reclassified as intermediate polars. 

 

\section*{Acknowledgments}
We thank the anonymous referee and
Chris Mauche for helpful comments. We also thank the AAVSO for providing 
optical lightcurves.  These are based on observations by 
variable star observers worldwide. 
This research made use of data obtained from the 
Leicester Database and Archive Service (LEDAS) at the University of
Leicester. DSB acknowledges support of a PPARC studentship
and a two-month Monbusho fellowship. Astrophysics at the University of
Leicester is also supported through PPARC rolling grants.

\bibliographystyle{mnras3}
\bibliography{refs}

\end{document}